\documentclass[twocolumn,showpacs,preprintnumbers,footinbib]{revtex4}

	\usepackage{bm,enumerate,dcolumn,graphicx}
	\usepackage{color,amsmath,amssymb}
	\usepackage{fancyhdr}
\begin{document}

	\title{Drastically enhanced high-order harmonic generation from endofullerenes}
	%
	\author{T. Topcu$^1$} 
	\altaffiliation{Current address: Department of Mathematics, Virginia Tech, Blacksburg, Virginia 24061, USA }
	\email[]{turkert@vt.edu}
	\author{E. A. Bleda$^2$, and Z. Altun$^2$} 
	\affiliation{
	$^1$ Department of Mathematics and Statistics, University of Nevada, Reno, Nevada 89557, USA \\
	$^2$ Department of Physics, Marmara University, Goztepe Campus, Istanbul 34722, TURKEY 
	}

	\date{\today}
	
	\begin{abstract}
	
	Dynamically rich nature of the high-order harmonic generation process lends itself to a variety of ways to increase photon yield and extend the harmonic cut-off frequency. We show here that high-harmonic generation from an atom confined inside an attractive potential shell can show a dramatic increase in the photon yield in certain cases. We consider an endohedrally confined hydrogen atom inside a C$_{60}$ cage as an example, and consider three distinct physical situations in which the initial state is (1) entirely confined inside the C$_{60}$, (2) partially outside, and (3) mainly localized on the cage wall. We demonstrate that when the atom-cage system starts in a state with a classical turning point outside the C$_{60}$ shell, the high-harmonic photon yield can be enhanced up to 4 orders of magnitude when compared with a free atom in the same initial state. We explain the underlying physical mechanisms in each case using fully three-dimensional quantum simulations. This gives a prime example of how directly coupling an atom to a nanostructure can alter strong field processes in atoms in interesting ways. 
	
	\end{abstract}
		
	\maketitle
	
	\section{Introduction}
	
	High-order harmonic generation (HHG) is a non-linear process in which an atom or a molecule interacts with an intense laser field and emits a broadband spectrum of photons. The photons are coherent harmonics of the driving laser field and have a relatively uniform intensity distribution forming a plateau. Similar intensities of the high-harmonics in the plateau region allow scientists to synthesize intense attosecond pulses of sub-femtosecond durations extending from the extreme ultraviolet to the soft X-ray region.  These spatially and temporally coherent pulses give rise to new techniques for understanding the underlying dynamics of physical and chemical changes that occur in atomic and molecular systems at femtosecond and attosecond timescales~\cite{imaging1Lein2007, imaging2Hentschel2001, Yang2015, procMcGrath2014}. 
    
    Composing such short pulses in a meaningful way requires the high-harmonic photons to also have sufficiently large intensities.  Increasing emitted  photon intensities while pushing the harmonic cut-off to higher frequencies has therefore drawn effort. Such  techniques range from invoking many-electron effects~\cite{PaSa13, ShScTr11} to exploiting phase matching in macroscopic targets~\cite{GaSaCo99, SaAnPh98, AnMiDe97, TaKaIs07}. Another method exploits this by using nanometer size metal tips to increase the electric field felt by tunneled electrons to favor of the recombination process over ionization. A nanostructure is used to enhance the HHG yield in this case, although the atom itself is not directly coupled to it~\cite{plasmonHusakou2011, plasmonKim2008, inhomogYavuz2012, inhomogCiappina2012}. In this paper, we take a first step in this direction by directly coupling an atom to a nanostructure to alter the tunneling and recombination dynamics. We investigate possible ways in which the highly nonlinear HHG process can be modified to give increased photon yield to aid in the production of higher intensity pulses. 
	
	Understanding the basic dynamics of HHG will make it easier to see how we manipulate it to our advantage. The physical mechanism behind HHG can be modeled as a three-step process~\cite{Corkum1993}: (1) the electric field of the laser suppresses the Coulomb potential which allows the electron to tunnel into the continuum. (2) The laser field accelerates the free electron. When the sign of the electric field changes, the electron accelerates towards the parent ion. (3) Finally, the electron recombines with the parent ion and emits a photon with an energy that is equal to the sum of the binding energy and the kinetic energy gained during the propagation step (2). By coupling the atom to a fullerene cage, we will demonstrate that the dynamics in all these steps can be modified in interesting, and in some cases, useful ways.

	\begin{figure}[h!tb]
		\begin{center}
			\resizebox{0.9\columnwidth}{!}{\includegraphics[angle=0]{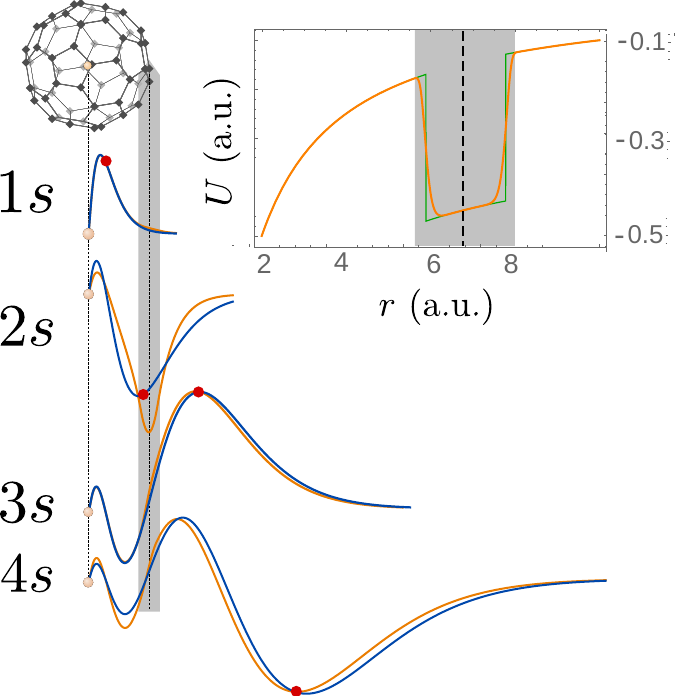}}
		\end{center}
		\caption{(Color online) Relative size scales involved in the harmonic generation from Ry atoms confined inside a C$_{60}$ molecule. The model potential commonly used to describe a C$_{60}$ molecule is a spherical potential shell with a radius of 6.84 a.u. The panel on the right shows the combined Coulomb potential and the  model potential for the C$_{60}$ molecule (solid green). In our simulations, we use the model seen as the solid orange curve. The eigenstates listed compare free atom (solid blue) and confined atom (solid orange) states, which we use as initial states in our simulations. The red points on the free atom eigenstates indicate $\langle r \rangle = 3n^2/2$.  
		}
		\label{fig:diagram}
	\end{figure}
	
	Here, we investigate high-order harmonic generation (HHG) from a hydrogen atom confined inside a fullerene, specifically, C$_{60}$. We solve the time-dependent Sch\"odinger equation (TDSE) within a single active-electron model to investigate three physically distinct situations: (1) when the atom is initially prepared in  the ground state which is entirely confined inside the fullerene cage, (2) in the 2s$^*$ state of the combined atom-fullerene system, in which the expectation value $\langle r\rangle$ falls inside the cage wall, and finally (3) in the 3s$^*$ state with the classical turning point outside the fullerene wall. 
	
	There are two important length scales in the problem: the expectation value $\langle r\rangle$ at $3n^2/2$, and the classical turning point at $2n^2$. When the atom starts in the ground state, we find that confinement introduces some enhancement in the photon yield compared to a free atom. This enhancement is lost when we increase the laser intensity. Initially preparing the combined atom-fullerene system in the first excited state (2s$^*$) results in a dramatically reduced cut-off frequency, which although undesirable, exhibits interesting physics. In this case, the electron is mainly localized on the C$_{60}$ cage, and we elucidate the physical mechanism at play using avoided crossings. Finally, we show that by preparing the atom in an excited state with a classical turning point outside the fullerene shell, the photon yield can be dramatically enhanced while retaining the same cut-off frequency obtained from a free atom. We find that this results from an afocal lensing effect, which is mediated by the spherically symmetrical fullerene cage.  We present momentum- and spatial-distributions of the time-dependent wave function to illustrate this mechanism.  

	The paper is organized as follows: we start by giving a brief account of our three-dimensional time-dependent simulations and the model we use to describe the atom-fullerene system in the laser field. In Sec.~\ref{sec:results}, we present results from our simulations and explain the underlying physical mechanisms based on our time-dependent simulations (Sec.\ref{subsec:h-1s}, and~\ref{subsec:h-3s}) and in terms of avoided crossings (Sec.~\ref{subsec:h-2s}). We conclude in Section~\ref{sec:conclusion} by suggesting possible future directions. 
	We label the bound states of the confined atom with an asterisk throughout the paper, {\it i.e.} 1s$^*$, 2s$^*$, 3s$^*$, whereas corresponding atomic states are labeled 1s, 2s, and 3s to differentiate the bound states of the H atom from those of the atom-fullerene system. We use atomic units unless we explicitly indicate otherwise. 
	
	\section{Simulations}\label{sec:theory}

   All of our simulations are based on {\it ab initio} solutions of the three-dimensional time-dependent Schr\"odinger equation (TDSE) in the length gauge. Detailed accounts of our simulations can be found in~\cite{TopRob12} and ~\cite{BleYavAlt13}. Here we only give a concise description of our methods emphasizing details most pertinent to present calculations. We start with numerical solutions of the TDSE. We then describe how we model the fullerene cage and describe a Green's function method for visualizing the wave function in space. We also perform calculations of the momentum space wave function to gain further insight into the dynamics, which we describe in Sec.~\ref{sec:theory:pmap}. 
   
   \subsection{Fully three-dimensional TDSE} \label{sec:theory:tdse}
   We carry out three-dimensional quantum calculations by solving the time-dependent Schr\"odinger equation~\cite{topcu07}. We decompose the time-dependent wave function in spherical harmonics $Y_{\ell,m}(\theta,\phi)$ as 
	\begin{equation}\label{scheq3d:decomp}
	\Psi(\vec{r},t)=\sum_{\ell=0}^{L_{\rm{m}}} f_\ell(r,t) Y_{\ell,m}(\theta,\phi) \;.
	\end{equation}
    The time-dependence is captured in the coefficient $f_\ell(r,t)$ discretized on a square-root mesh in the $r$-direction for each angular momentum. This is ideal for describing Rydberg states because it places approximately the same number of radial points between the nodes of excited states. On this grid, the Schr\"odinger equation is 
    \begin{equation}\label{eq:scheq_prob} 
	\left[\text{i} \frac{\partial}{\partial t} - H(r,l,t) \right] \psi(r,l,t) = 0 \;.
	\end{equation}
    We split the total hamiltonian into an atomic hamiltonian plus the interaction hamiltonian: $H(r,l,t)=H_\text{A}(r,l) + H_\text{L}(r,t)-E_\text{0}$, where we subtract the initial state energy from the total hamiltonian to reduce the numerical phase accumulation over time. The atomic hamiltonian $H_\text{A}$ and the hamiltonian describing the interaction of the atom with the laser field in the length gauge are 
	\begin{eqnarray}\label{scheq3d:split}
	H_\text{A}(r,l) &=& -\frac{1}{2}\frac{\text{d}^2}{\text{d}r^2}-\frac{1}{r}
		+\frac{l(l+1)}{2r^2}\\
	H_\text{L}(r,t) &=& F(t) z \cos(\omega t) \;.
	\end{eqnarray}
	We use the lowest order split operator technique to evolve the total wavefunction  in time according to~\eqref{eq:scheq_prob}, where each split piece is propagated using an $\mathcal{O}(\delta t^3)$ implicit scheme, which is exactly unitary. A detailed account of the $\mathcal{O}(\delta t^3)$ implicit method and the split operator technique we use is given in~\cite{topcu07}. 
	
	We report square of the Fourier transform of the dipole acceleration $\langle\ddot z\rangle (t)$, $|a(\omega)|^2$, since the radiated power in the length gauge is proportional to it: $S(\omega)=2\omega^4 |a(\omega)|^2/(3\pi c^3)$. We also calculate the dipole and the velocity forms of the spectra, $|d(\omega)|^2 = \omega^4 |\langle z\rangle (\omega)|^2$ and $|v(\omega)|^2=\omega^2 |\langle\dot z\rangle (\omega)|^2$ for comparison. The level of agreement between these forms gives us an idea about how weak the laser pulse is since $|d(\omega)|^2$, $|v(\omega)|^2$, and $|a(\omega)|^2$ agree well when the pulse is effectively weak. These different forms start to differ in the strong field regime when the final state of the atom is strongly mixed. We find that all three forms of the spectra agree very well throughout the plateau region in all our calculations except when we start in the 3s$^*$ state. In this case, the laser pulse strongly mixes adjacent $n$-manifolds, and leaves the confined system in a superposition state. The time-dependent dipole acceleration is given by 
	\begin{equation}\label{scheq3d:decomp}
	\langle\ddot z\rangle (t) = -\langle \psi(r,t)| [H,[H,z]] |\psi(r,t)  \rangle \;.
	\end{equation}

	We keep track of the time-dependent ionization probability, $P(t)$, as well, and the total probability amplitude inside the box not bound to an eigenstate of the atom-fullerene system, $\widetilde{P}(t)$:  
	\begin{eqnarray}\label{eq:ion1} 
	P(t) &=& 1- \smallint\limits_{0}^{R} {{{\left| \psi (r,t) \right|}^{2}}dr} \;, \\
	\widetilde{P}(t) &=& \smallint\limits_{0}^{R} {{{\left| \psi (r,t) \right|}^{2}}dr} 
 		- \sum_{n,\ell} \left| \smallint\limits_{0}^{R} \phi_{n,\ell}(r) \psi(r,t) dr \right|^2 \;. 
	\end{eqnarray} 
	Here $R$ is the radial extent of the box, and $\phi_{n,\ell}(r)$ are the eigenstates of $H_{\rm{A}}$. $\widetilde{P}(t)$ provides us with a measure of slowly escaping amplitude that is still inside the box after the laser pulse. Note that $\widetilde{P}(t) \rightarrow 0$ in the long-time limit because the sum in~\eqref{eq:ion1} is performed over a complete set of states inside the box. 
	
	We choose the intensity and wavelength in our simulations to prevent structural changes to the C$_{60}$ molecule. We use an 800 nm laser at an intensity of $5\times 10^{13}$ W/cm$^2$ for the simulations of HHG out of the ground state. The pulse duration, in this case, is 4-cycles, corresponding to a $\sim$10 fs pulse at FWHM; and the intensity is below the saturation intensity of the first charge stage of C$_{60}$ at 800 nm~\cite{BecBecFai06}. On the other hand, multiphoton ionization and fragmentation processes can still take place in intensities below the saturation limit, beyond which the fullerene molecules are completely destroyed, and a hot carbon plasma is created. In this regime, ionization stage  rarely goes beyond C$_{60}^{2+}$ due to competition between the C$_{60}$ lattice modes and electron emission, and fragmentation takes place by emission of C$_{60-2m}^{q+}$ fragments where $m=0,1,\cdots$ and $q=1,2,3$~\cite{HunStaHui96} in the intensity range $10^{13}$ - $10^{14}$ W/cm$^2$ at 790 nm. In this case, fragmentation takes place via multiphoton excitation of a giant plasmon resonance at $\sim$20 eV and is a perturbative process at these intensities ~\cite{HunStaHui96}. 
    
    Another physical process we ignore is HHG from the C$_{60}$ molecule itself~\cite{CiaBecBec08}. The overall dielectric response of the C$_{60}$ molecule is dominated by the collective motion of the $\pi$-electrons and the large electron density relative to the Debye length leads to efficient screening of the laser field inside the molecule~\cite{HunStaHui96}. Fields even stronger than what we consider here would be hardly felt by the electrons in the interior of the molecule, and no efficient tunneling takes place from the C$_{60}$ molecule. The fullerene will ionize and fragment long before tunneling can take place to start an efficient HHG process. 
    
    When simulating HHG from excited states, we scale the wavelength and the intensity as $800 n^3$ nm and $(1.5/n^8)\times10^{14}$ W/cm$^2$ to keep the time-dependent dynamics in the same physical regime throughout our simulations involving Ry states. These laser frequencies and intensities are far below those that would cause structural changes in the C$_{60}$ molecule for $n\ge 2$. 
	
	\subsection{The model potential}\label{sec:theory:c60}
	
	We place the H atom at the center of a spherically symmetric potential shell centered at the position of the atomic nucleus ($r=0$). In our simulations, we smooth out the sharp edges of this model potential to prevent artificial high-frequency components from appearing in the harmonic spectra (solid orange curve in Fig.~\ref{fig:diagram}). We model the C$_{60}$ cage with 
	\begin{equation} \label{eq:ourc60}
	 U_{{\rm C}_{60}}(r) = U_0 \exp\left (-\frac{(r-r_{\rm c})^{10}}{a} \right ) \;,
	\end{equation}
	where $U_0$ is the depth and $r_c=6.83$ is the distance from the nucleus to the middle of the shell wall. The parameter $a$ is associated with the width of the cage. We find $a$ by least square fitting ~\eqref{eq:ourc60} to the following radial square-well potential for the $C_{60}$ structure~\cite{DolCraGul09}: 
	\begin{equation}\label{eq:dolmatovc60}
	   U_{{\rm C}_{60}}(r) = \left\{
		 \begin{array}{c c}
			-0.3 & 5.89\leq r\leq 7.78 \\
			 0   & {\rm otherwise} 
		 \end{array}
	   \right .
	\end{equation} 
	and determine it to be 0.83 (a.u.)$^{10}$. We repeated some of our calculations using the model potential in Eq.~\eqref{eq:dolmatovc60}, and found no significant difference from the harmonic spectra generated using~\eqref{eq:ourc60}. 

	We also want to point out that the atoms inside real endofullerenes are not centered as they are in our model, but are off-centered, and typically situated near the inner wall of the fullerene. Accurately reflecting this would affect our results. However, a real experiment is performed on a macroscopic ensemble of atoms, and a uniform distribution of positions of atoms inside fullerenes would produce a macroscopic response that is closely mimicked by a response obtained from our model simulations where the atom is assumed to be centered. This is because, in the absence of substantial non-linearities, the macroscopic response can be qualitatively reproduced by averaging single atom responses from atoms inside the gas. This is why, for example, numerical data for photoionization of endofullerenes using the model we employ here agree so well with data from experiments in the literature (see, e.g.,~\cite{MuScPh07, ChPhMs10, DolCraGul09}). Although the calculations assume that the atom is centered at the origin, the experiment is performed on a gas target which contains a large number of atoms, and that what is measured is the macroscopic response. 
	
	\subsection{Green's function method}\label{sec:theory:greens}
    We demonstrate the physical mechanism behind the enhancement when the atom-fullerene system starts in the 3s$^*$@C$_{60}$ state by observing the behavior of the probability amplitude near the fullerene cage. Discerning details of the interesting dynamics of the wave function near the fullerene wall is, however, difficult using the fully three-dimensional simulations we describe in Sec.~\ref{sec:theory:tdse}. The reason is that most of the initial wavefunction remains unperturbed by the driving laser field, and the part of the wavefunction that takes part in the interesting dynamics which generate high-harmonic photons is buried underneath the initial wavefunction, making it difficult to see. To observe the behavior of the wavefunction that contributes to the HHG process, we, therefore, employ a time-dependent method similar to the first-order time-dependent perturbation theory which we briefly describe below. Details of this method can be found, {\it e.g.}, in~\cite{TopRob12}. 
    
    In this method, we express the total wavefunction as a superposition of the initial state, and the time-dependent correction, $\psi(r,l,t)=\psi_0(r,l)+\psi_1(r,l,t)$, then write the time-dependent Schr\"odinger equation as 
	\begin{equation}\label{scheq_rate}
	\left[\text{i} \frac{\partial}{\partial t} - H(r,l,t) \right] \psi_1(r,l,t) 
	= H_\text{L}\psi_0 \;.
	\end{equation}
	Here $H_\text{L}(r,t)\psi_0(r,l)$ acts as a source term, and the interesting dynamics is encoded in $\psi_1(r,l,t)$. This method is particularly useful when the amplitude in $\psi_1(t)$ is much smaller than the amplitude in $\psi_0$ as is the case in our simulations. The wave function $\psi_1(t)$ is initially zero everywhere, and it encodes the time-dependent corrections to the unperturbed wave function $\psi_0$. This method allows for atomic processes of all orders, such as tunneling, as well as single- and multi-photon processes.

	\subsection{Momentum distribution}\label{sec:theory:pmap}
	As another way of assessing the physical mechanism of the enhancement when the atom starts in an excited state, we also observe the time-evolution of the momentum distribution. We evaluate the momentum distribution following the procedure outlined in Ref.~\cite{vrakking} with the exception that it is the momentum distribution of the total wavefunction and not just the escaping amplitude. For the sake of completeness, we briefly describe the method here. During the time evolution of Eq.~\eqref{eq:scheq_prob}, the ionized part of the wave function is removed from the box after every time step to prevent unphysical reflections from the radial box edge. This is done using a mask function $\mathcal{M}(r)$, spanning the final 1/3 of the radial box. The removed part of the wavefunction is 
	\begin{equation}\label{scheq3d:masked}
   	\Delta \psi_{l}(r,t')= [1-\mathcal{M}(r)]\;\psi_{l}(r,t') \;,
	\end{equation}
    and we evaluate $\Psi_{l}(r,t')=\psi_{l}(r,t')+\Delta \psi_{l}(r,t')$. In order to emphasize the momentum distribution beyond the peak of the depressed Coulomb potential beyond $r>1/\sqrt{F_{\text{\rm p}}}$, we apply a mask function to $\Psi_{l}(r,t')$ in the region $r<1/\sqrt{F_{\text{\rm p}}}$ which exponentially suppresses the wavefunction as $r\rightarrow 0$.  Here $F_{\text{\rm p}}$ is the peak electric field of the laser pulse. We then Fourier transform $\Psi_{l}(r,t')$ to obtain the momentum space wave function $\phi(p_{\rho},p_z,t')$, 
	\begin{eqnarray}\label{scheq3d:pmap}
	   \phi(p_{\rho},p_z,t') = &&2 \sum_{l}(-i)^{l}
	      \;Y_{l,m}(\theta,\varphi) \nonumber \\ 
		   &&\times\int_0^{\infty} j_{l}(pr) \Psi_{l}(r,t') r^2 \;dr \;.
	\end{eqnarray}
    Here the momentum is $p=(p_{\rho}^2+p_z^2)^{1/2}$ and $j_{l}(pr)$ are the spherical Bessel functions. Because the laser field is polarized along the $z$-axis, $p_\rho$ is the perpendicular component of the wavepacket $p_{\perp}$ to the laser polarization, whereas $p_{||}$ is along $p_z$.

	\section{Results}\label{sec:results}
	
	We consider the following three distinct situations: (1) when the initial state of the combined atom-fullerene system is entirely confined inside the cage (H(1s$^*$)@C$_{60}$), (2)  when $\langle r \rangle=3n^2/2$ falls in the fullerene wall (H(2s$^*$)@C$_{60}$), and (3) when the classical turning point at $2n^2$ is outside the cage wall ((H(3s$^*$)@C$_{60}$)). In the 1s$^*$ state, the classical turning point is well inside the cage. We observe $\sim$4 orders of magnitude enhancement in yield in case (3). In contrast, we see a drastic reduction in cut-off in case (2) -- although not desirable for applications -- displays interesting physics due to the existence of avoided crossings when we vary the cage depth. 
	
	\subsection{H(1s$^*$)@C$_{60}$}\label{subsec:h-1s}
\begin{figure}[h!tb]
		\begin{center}
			\resizebox{1.0\columnwidth}{!}{\includegraphics[angle=0]{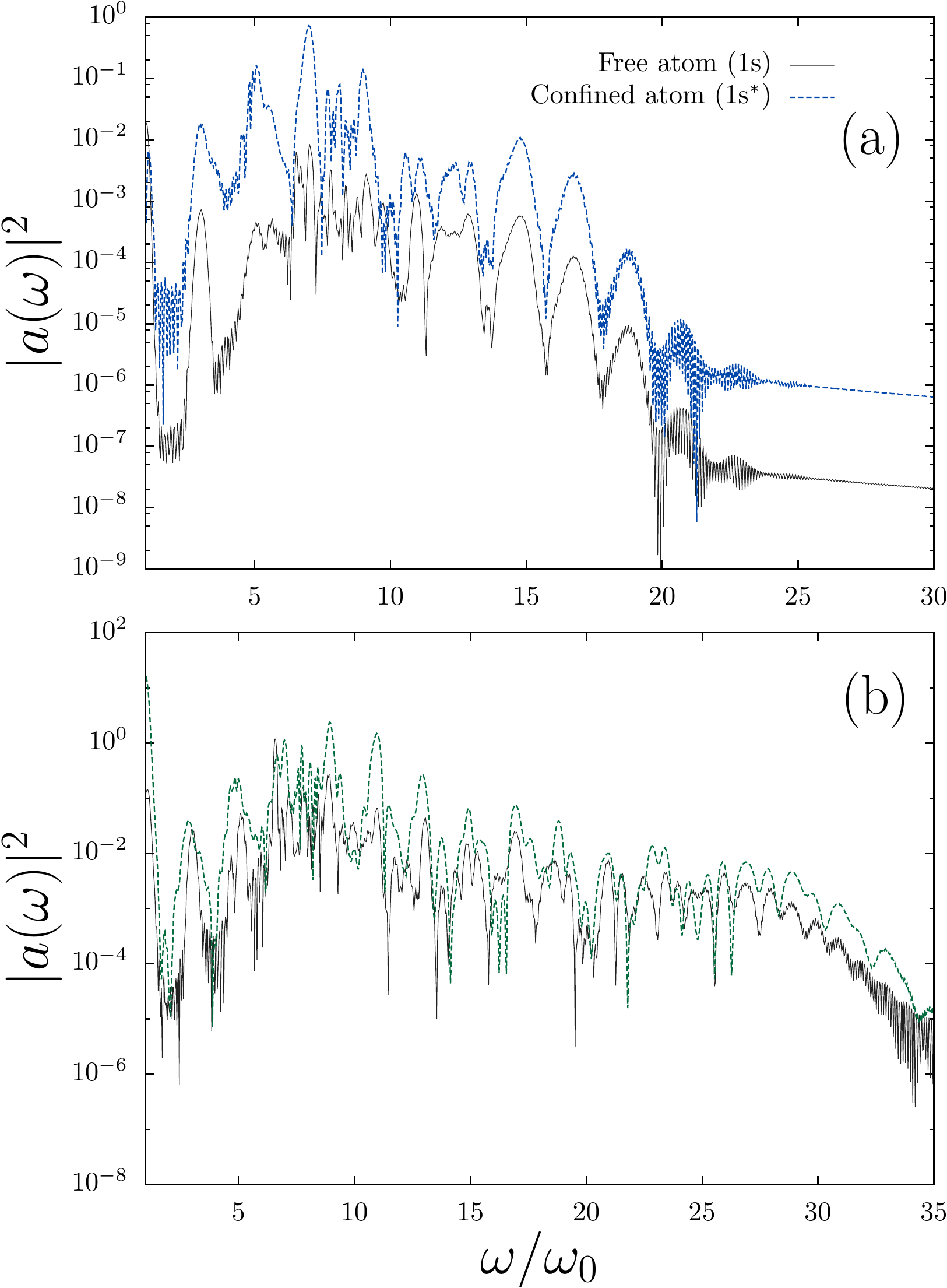}}
		\end{center} 
		\caption{(Color online) Panel (a) compares HHG spectra from free (solid black) and confined (dashed blue) H atoms initially prepared in the ground states. The laser intensity is $5.0\times 10^{13}$ W/cm$^2$ and the wavelength is 800 nm. In this case, yield from the endofullerene is roughly an order of magnitude larger than the yield from the H atom alone. This enhancement, however, goes away when the intensity is increased, which is shown in the lower panel in which the laser intensity is $1.5\times 10^{14}$ W/cm$^2$. 
		}
		\label{fig:hhg-h1s}
	\end{figure}

    Results in Fig.~\ref{fig:hhg-h1s}(a) compare HHG spectra from H and H@C$_{60}$ when the atom and the combined atom-fullerene system is initially prepared in their respective ground states, 1s and 1s$^*$. The laser intensity and  wavelength in this case are $5.0\times 10^{13}$ W/cm$^2$ and 800 nm. In the ground state of the endofullerene, $2n^2 < r_c$ and the radial wavefunction is entirely confined inside the fullerene cage. This results in bound state energies for the free and the confined systems being the same within one part in $10^5$. 
    
    Comparing spectra in the dipole, velocity and acceleration forms, we find that all gauges agree well throughout the plateau region until after the cut-off. Additionally, the ionization probability after the laser pulse is well below a percent for the free atom and $\sim$4\% for H(1s$^*$)@C$_{60}$. Low ionization probabilities combined with good agreement between different gauges in the plateau region suggests that we are in the weak and short pulse regime in Fig.\ref{fig:hhg-h1s}(a)~\cite{BanCheDie09,BleYavAlt13}. Furthermore, macroscopic effects such as depletion and dispersion due to free electrons from the ionized atoms would be negligible. 
    
	The enhancement in Fig.~\ref{fig:hhg-h1s}(a) goes away when we increase the intensity keeping wavelength the same in Fig.~\ref{fig:hhg-h1s}(b). In this case, the intensity is three times larger than that in Fig.~\ref{fig:hhg-h1s}(a) at $1.5\times 10^{14}$ W/cm$^2$. The enhancement in Fig.~\ref{fig:hhg-h1s}(a) when $I=5.0\times 10^{13}$ W/cm$^2$ can be understood using Fig.~\ref{fig:hhg-h1s-pot} below. 

	Fig.~\ref{fig:hhg-h1s-pot} shows potentials for the H@C$_{60}$ system depressed by the laser field at two different peak intensities: $5\times 10^{13}$ W/cm$^2$ (dashed blue curve), and $1.5\times 10^{14}$ W/cm$^2$ (green dash-dotted curve).  We also show the effective potential in the absence of laser field (solid red curve). The dotted horizontal line marks the energy of the ground state, H(1s$^*$)@C$_{60}$. The tunneling distance through the effective potential when $I=5\times 10^{13}$ W/cm$^2$ is less than the tunneling distance through the corresponding cage free depressed potential at the same peak intensity. Note that the potentials in Fig.~\ref{fig:hhg-h1s-pot} have the same shape as their free-atom counterparts which lack the potential well centered at $r_c$. A shortened tunneling distance translates into a larger tunneling probability, which in turn results in a larger recombination probability.  On the other hand, the tunneling distance when $I=1.5\times 10^{14}$ W/cm$^2$ is essentially the same as the tunneling distance through the corresponding {\it cage-free} depressed potential. When the laser intensity is higher, the presence of the potential well describing the C$_{60}$ potential becomes less important because the potential barrier is further suppressed. This is why the intensity of the emitted harmonics is larger for H@C$_{60}$ than those generated from the H atom in Fig.~\ref{fig:hhg-h1s}(a), whereas they are similar for the larger laser intensity in Fig.~\ref{fig:hhg-h1s}(b). 

	\begin{figure}[h!tb]
		\begin{center}
			\resizebox{1.0\columnwidth}{!}{\includegraphics[angle=0]{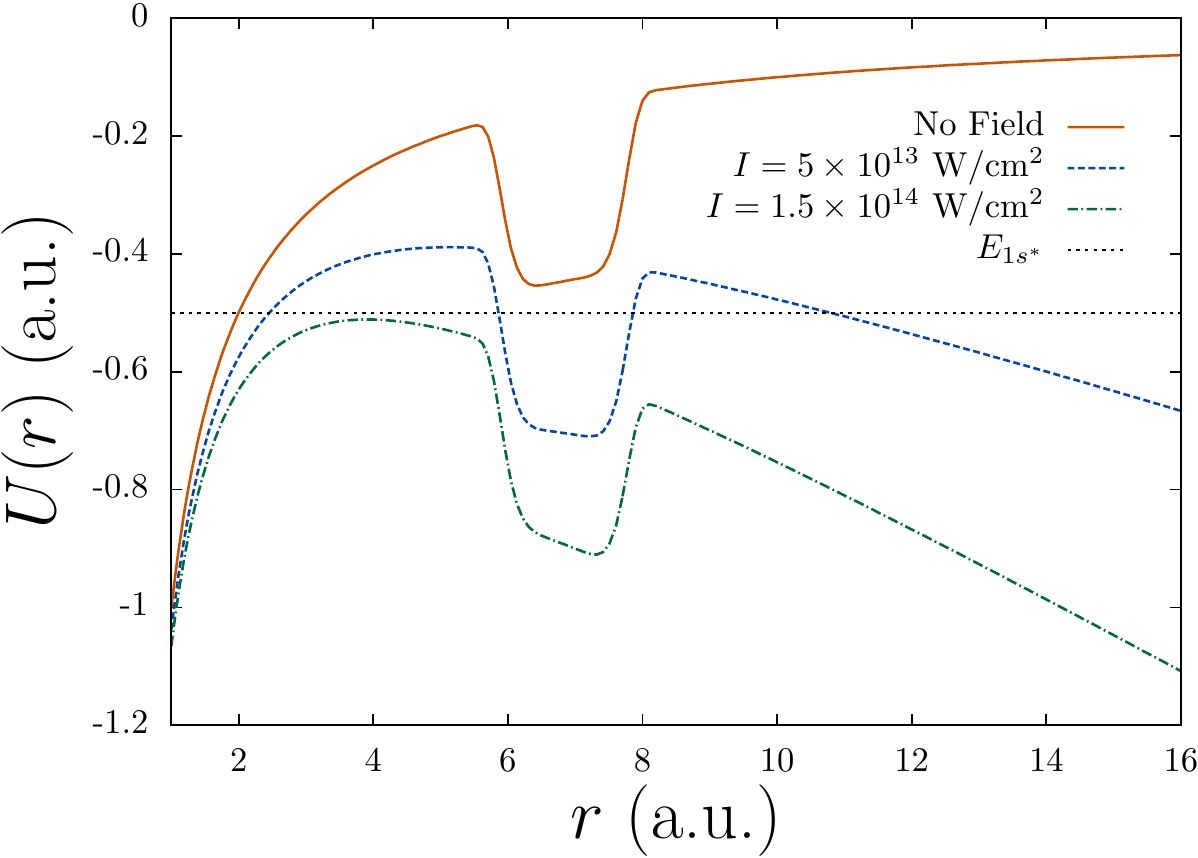}}
		\end{center} 
		\caption{(Color online) Combined Coulomb and model fullerene potential depressed by the laser field at the peak of the laser pulse for the two intensities in Fig~\ref{fig:hhg-h1s}. For reference, zero field potential is also shown. The potentials do not become depressed more than shown in the figure at any time during the pulse. 
		}
		\label{fig:hhg-h1s-pot}
	\end{figure}
	
\begin{figure}[h!tb]
		\begin{center}
			\resizebox{1.0\columnwidth}{!}{\includegraphics[angle=0]{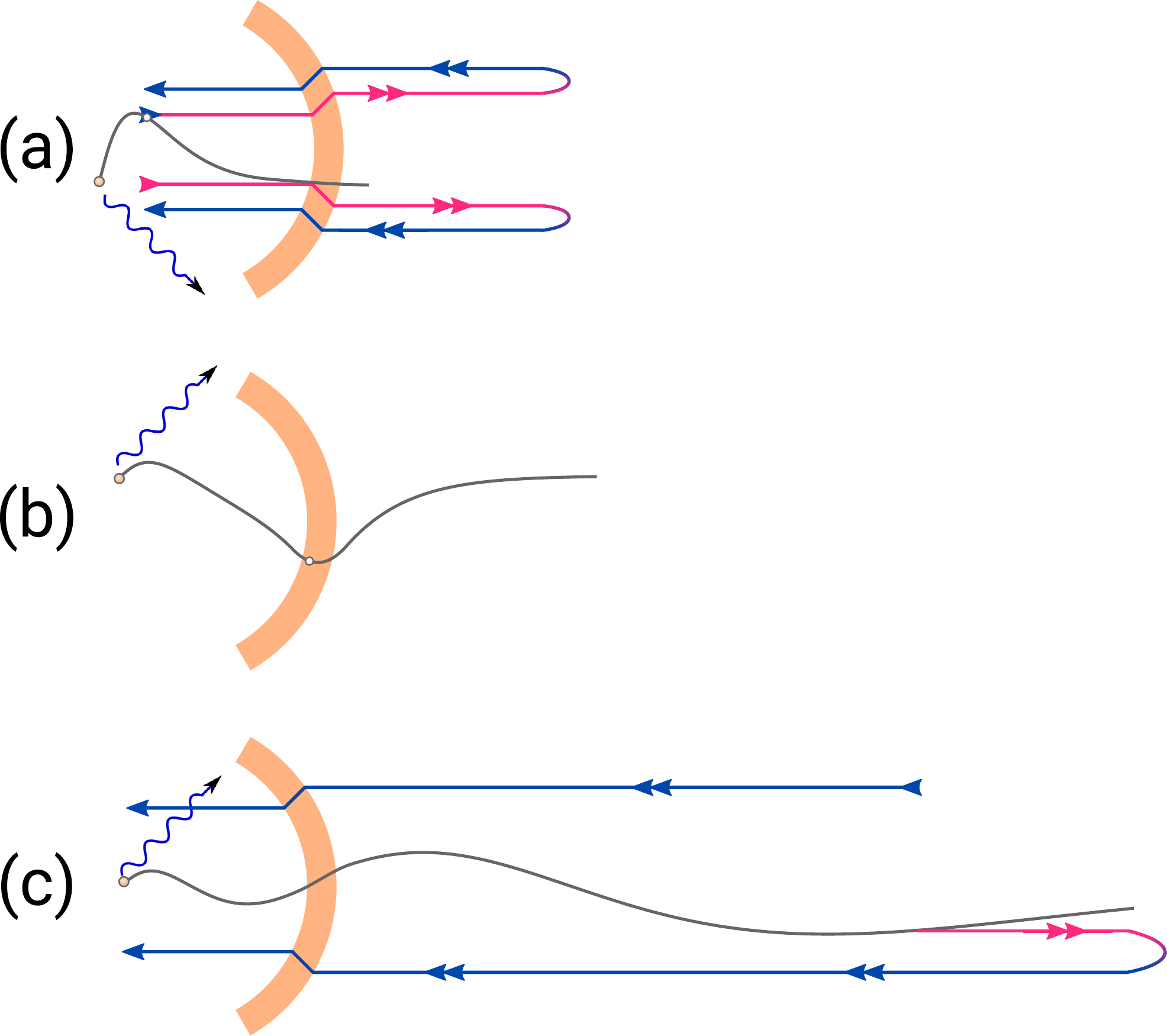}}
		\end{center}
		\caption{(Color online) The afocal lensing effect leading to three different situations when H@C$_{60}$ is initially prepared in:  (a) the ground state where both tunneling and recombination takes place inside the fullerene cage, (b) the 2s$^*$ state when electron is mainly localized on the fullerene wall, and (c) the 3s$^*$ state where the tunneling happens outside the cage but the recombination takes place inside. 
		}
		\label{fig:3cases}
    \end{figure}

	The potential well describing the C$_{60}$ shell seen in Fig.~\ref{fig:hhg-h1s-pot} acts like an afocal lens (also called a zero meniscus lens) for both the outgoing (tunneling) and the incoming (recombining) wave packets at the first and third steps of the HHG process. Such a lens has an infinite focal length, meaning it neither focuses nor diverges the beams. However, it squeezes or spreads them in space in a direction perpendicular to their direction of propagation (Fig.~\ref{fig:3cases}). If the rays are incoming (Fig.~\ref{fig:3cases}(a) and (c)), they are squeezed together; and if they are outgoing (reverse the directions of the rays in Fig.~\ref{fig:3cases}), they spread out, without changing direction. 
	
	For HHG out of the 1s$^*$ ground state, the classical turning point is at $2n^2=2$ a.u., which is inside the C$_{60}$ cage. Therefore, the tunneling happens inside and the outgoing wavepacket spreads out as it exits the cage. After it is turned around and driven back to the atom for recombination, the incoming wavepacket is squeezed together again before it recombines inside the cage (Fig.~\ref{fig:3cases}(a)). As a result, both the outgoing and the incoming wavepackets have the same spread, and there is nothing different in either the tunneling or the recombination stages compared to a free atom. As a result, we see similar emission spectra in Fig.~\ref{fig:hhg-h1s}(b) for larger laser intensity, which renders the cage potential irrelevant as we discussed using Fig.~\ref{fig:hhg-h1s-pot} above. 

    There are two crucial points to emphasize here: Firstly, the tunneling takes place near the classical turning point at $2n^2$, and the recombination occurs near the nucleus, both of which are inside the fullerene shell for the 1s$^*$ state. The second crucial point is that the incoming wavepacket mainly recombines into the state out of which it had initially tunneled. 
	
	\subsection{H(2s$^*$)@C$_{60}$}\label{subsec:h-2s}
\begin{figure}[h!tb]
		\begin{center}
			\resizebox{1.0\columnwidth}{!}{\includegraphics[angle=0]{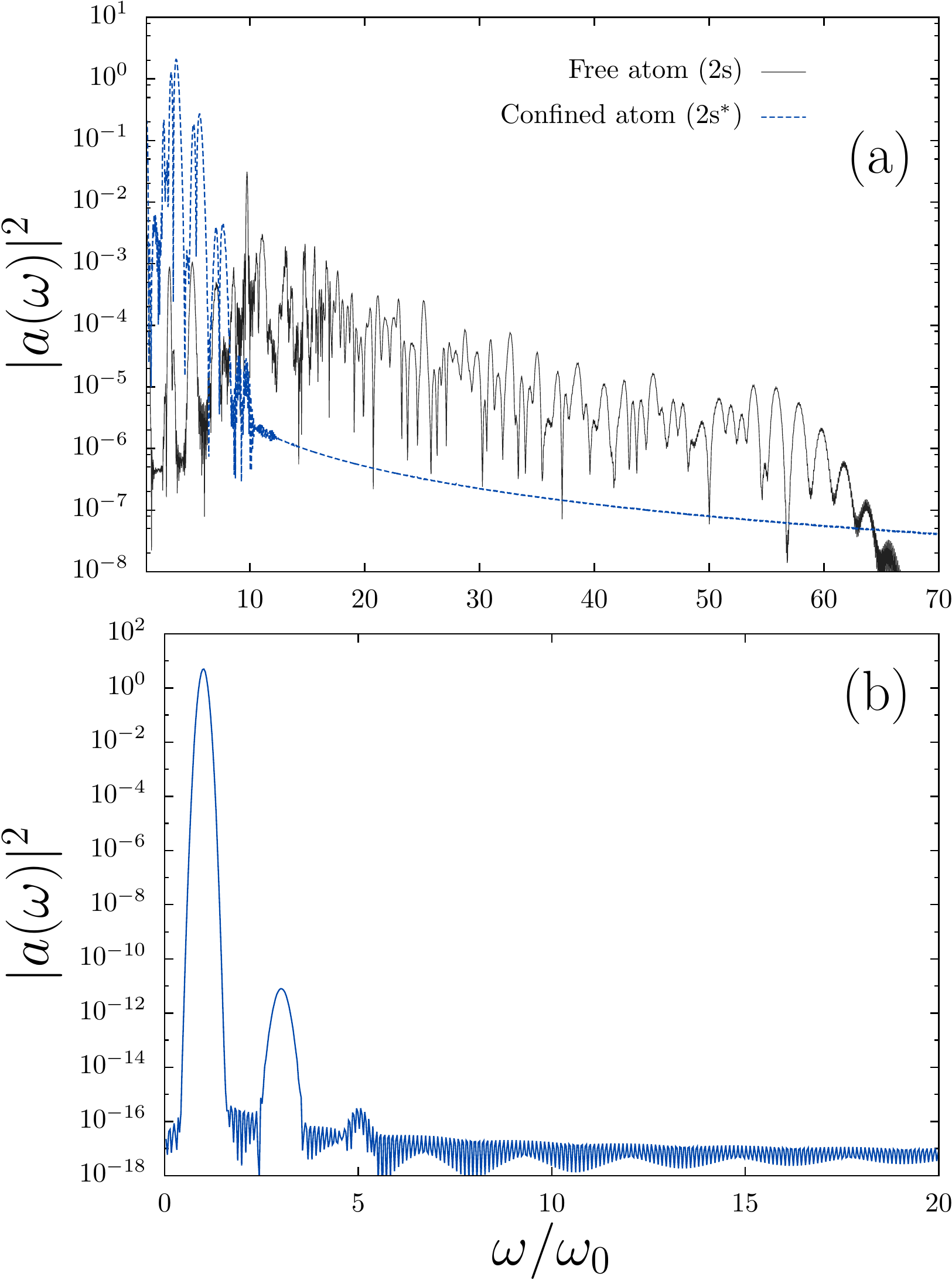}}
		\end{center}
		\caption{(Color online) Panel (a) compares spectra from a free H atom in the 2s state (black solid) with that from H@C$_{60}$ endofullerene initially prepared in the first excited state 2s$^*$ (blue dotted). For H@C$_{60}$, the cut-off is reduced by about a factor of 6 because the 2s$^*$state is radially localized on the fullerene cage wall. Panel (b) shows the spectrum from the ground state of the spherically symmetric shell potential (Eq.~\eqref{eq:ourc60}) with its depth adjusted so that its ground state has the same binding energy as the 2s$^*$ state of H@C$_{60}$. In these simulations, the laser intensity and wavelength are $5.0\times 10^{13}$/$n^8$ W/cm$^2$ and $800 n^3$ nm where $n=2$.  
		}
		\label{fig:hhg-h2s}
    \end{figure}

	The spectra in Fig.~\ref{fig:hhg-h2s}(a) are obtained when when the atom and the endofullerene are initially prepared in their first excited states, 2s and 2s$^*$. The laser intensity and the wavelength are $5.0\times 10^{13}$/$n^8$ W/cm$^2$ and $800 n^3$ nm where $n=2$. In the 2s$^*$ state of the H@C$_{60}$ system, $\langle r \rangle=(3/2)n^2 =6$ a.u. falls inside the wall of the fullerene cage (Eq.~\eqref{eq:ourc60}). As a result, the bound state energy of the 2s$^*$ state is  more deeply bound at -0.239 a.u. than the 2s state at -0.125 a.u. Starting in the 2s$^*$ state, we see a drastic reduction of the cut-off harmonic in the H@C$_{60}$ spectrum at $\omega/\omega_0$$\sim$10 compared to the H spectrum at $\omega/\omega_0$$\sim$54. Although a reduction in the cut-off frequency is not something practically desirable, we will show below that it follows from interesting physics. It also demonstrates that HHG spectrum from a confined atom can be turned on and off by changing the depth of the confining potential. 
    
    For both the free and the confined atoms, the dipole, velocity and acceleration forms of the emitted power agree well everywhere throughout the entire plateau region of the spectra for both the free and the confined species. Also, there is practically no ionization from either of the two systems by the end of the laser pulse. Lack of ionization means the reduction in the cut-off for the H@C$_{60}$ system does not follow from depletion. In fact, we see that $\omega/\omega_0 \lesssim 10$ is the region in which HHG spectrum from the spherically symmetric cage potential~\eqref{eq:ourc60} resides in the absence of the H atom when the fullerene model is prepared in its only bound state. This spectrum is shown in Fig.~\ref{fig:hhg-h2s}(b), where the depth of the C$_{60}$ shell is adjusted so that its ground state energy matches the energy of the H(2s$^*$)@C$_{60}$ system. Both spectra show strong harmonics below $\omega/\omega_0 \lesssim 5$, which can be understood within a picture based on avoided crossings as we explain below. 
	
\begin{figure}[h!tb]
            \begin{center}
                \resizebox{1.0\columnwidth}{!}{\includegraphics[angle=0]{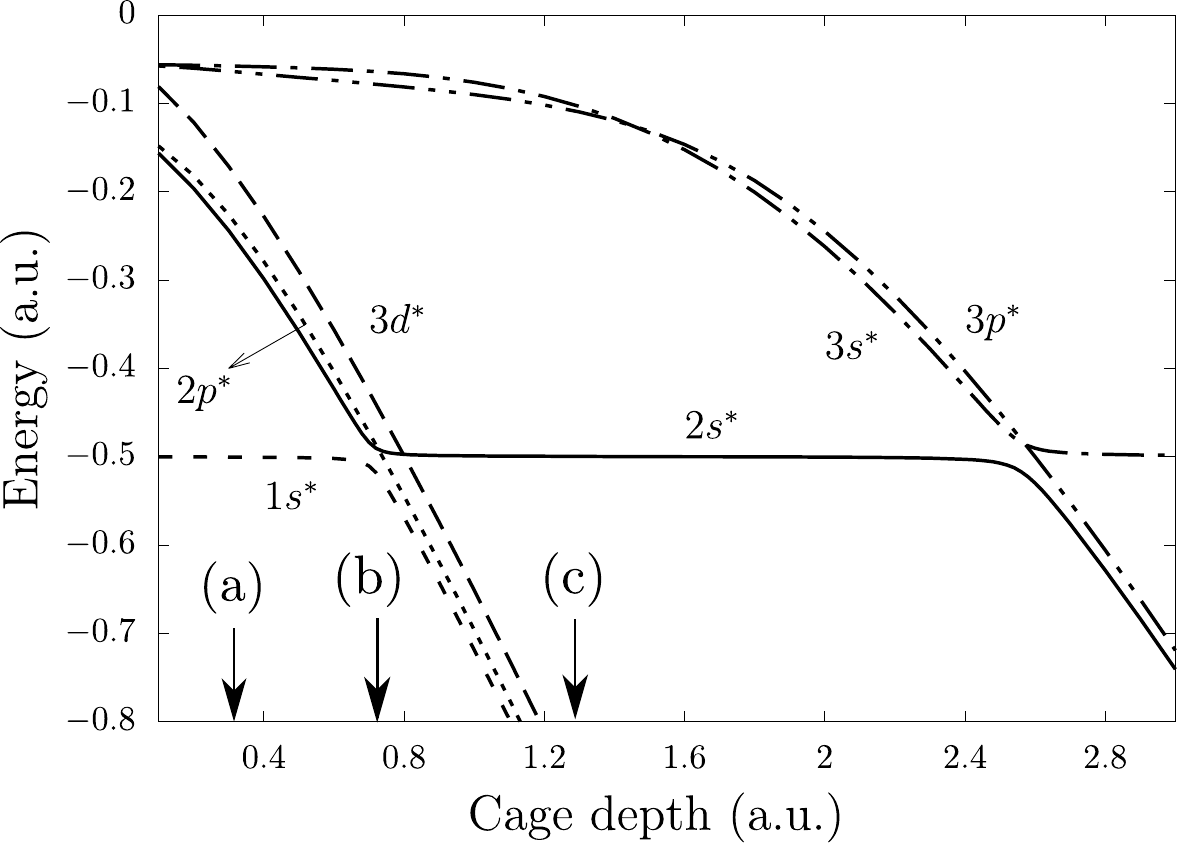}}
            \end{center}
            \caption{(Color online) Energy level curves for the 1s$^*$, 2s$^*$ and 3s$^*$ states of H@C$_{60}$ as a function of the cage depth $U_0$. There are avoided crossings at $U_0$$\sim$0.7 a.u. between the 1s$^*$ and the 2s$^*$ states, and at $U_0$$\sim$2.6 a.u. between the 2s$^*$ and the 3s$^*$ states. States with different parity do not avoid each other. Three points in cage depth are labeled at (a) 0.3 a.u., (b) 0.7 a.u., and (c) 1.3 a u.. 
        }
            \label{fig:enrCurves}
    \end{figure}

\begin{figure}[h!tb]
            \begin{center}
                \resizebox{1.0\columnwidth}{!}{\includegraphics[angle=0]{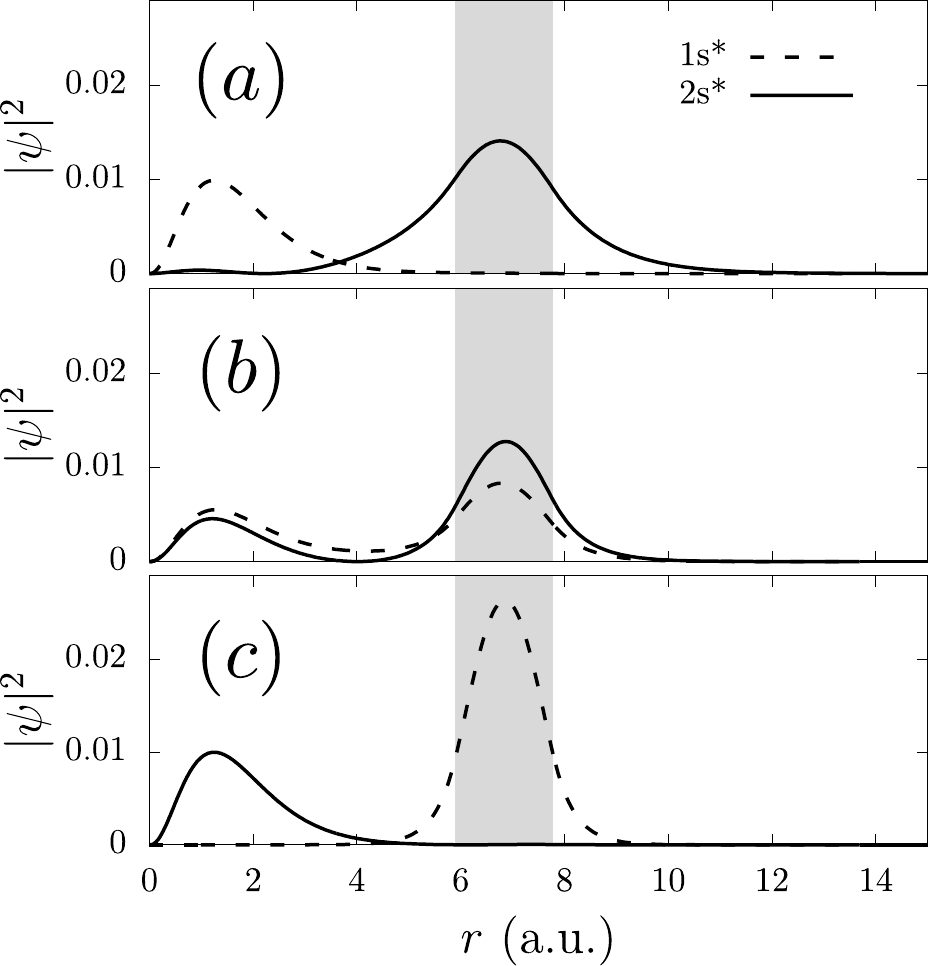}}
            \end{center}
            \caption{(Color online) Radial distributions of the 1s$^*$, 2s$^*$ states of H@C$_{60}$ at three cage depths (labeled a, b, c in Fig.\ref{fig:enrCurves}) around the first avoided crossing. The radial extent of the cage potential is highlighted. The cage depth is swept from top (a) to the bottom panel (b) in Fig.\ref{fig:enrCurves}. Before the avoided crossing, the 2s$^*$ state is mainly localized on the cage wall in panel (a), which moves to where the 1s$^*$ state was localized after the avoided crossing in (c). 
            }
            \label{fig:2s-wf}
    \end{figure}

	In the 2s$^*$ state of H@C$_{60}$, the wave function is mainly localized on the C$_{60}$ wall (Fig.~\ref{fig:3cases}(b)). As a result the depressed potential through which it needs to tunnel is effectively narrower compared to the situation in which the initial state is localized inside the C$_{60}$ wall (1s$^*$ state). Evolution of the H@C$_{60}$ energy levels with varying confining potential depth are presented in Fig.~\ref{fig:enrCurves}. Before the avoided crossing in Fig.~\ref{fig:enrCurves}, the 2s$^*$ state is localized on the fullerene wall, and can tunnel either to the left or the right of the potential barrier. This point is labeled (a) in Fig.~\ref{fig:enrCurves}, and the corresponding radial distributions of the states are shown in Fig.~\ref{fig:2s-wf}. In this case, the electron will tunnel to the left of the barrier towards the atomic core more preferentially, because of the asymmetric shape of the depressed Coulomb barrier in which it sits. This substantially reduces the tunneling amplitude for the electron wave packet that can effectively contribute to the HHG process, which is most efficient following a long excursion in the laser field before recombination. This accounts for the much shorter cut-off harmonic seen in Fig.~\ref{fig:hhg-h2s}(a) for the 2s$^*$ initial state  compared with the spectrum from the H atom starting in the 2s state. Note that the 2s$^*$ state is more deeply bound at $\sim$-0.24 a.u. than the 2s state at $\sim$-0.125 a.u., which, by itself, works against the factor of $\sim$6 reduction in cut-off frequency. 

	We will now analyze the reduction in cut-off seen in Fig.~\ref{fig:hhg-h2s}(a) using a simple 2-level picture to understand the behavior of the bound states of H@C$_{60}$. We examine the energy shifts induced by the confining potential based on the relative strengths of the matrix elements. 
    
    The total Hamiltonian is $H=H_{a}+U_{c}$, where $H_{a}$ and $U_{c}$ are the atomic hamiltonian and a spherically symmetric confining potential. The confining potential has the form~\eqref{eq:ourc60} in which we will vary its depth. Considering only the two lowest energy levels, the matrix representation of $H$ in the eigenbasis of $H_a$ is 
    \begin{equation}\label{eq:hmat}
        \begin{pmatrix}
            E_1  & 0 \\
            0  & E_2 
        \end{pmatrix}
        +
        \begin{pmatrix}
            W_1  & W \\
            W^*  & W_2 
        \end{pmatrix}
        =
        \begin{pmatrix}
            E_1+W_1  & W \\
            W^*  & E_2+W_2 
        \end{pmatrix} \;.
    \end{equation}

    Here the first term on the left represents the unperturbed atom whereas the second term is the confining potential. The matrix elements are: $E_1 = \langle 1s | H_a | 1s \rangle$, $E_2 = \langle 2s | H_a | 2s \rangle$, $W_1 = \langle 1s | U_c | 1s \rangle$, $W_2 = \langle 2s | U_c | 2s \rangle$, and $W = \langle 1s | U_c | 2s \rangle$. Diagonalizing this Hamiltonian results in the following eigenvalues: 
    \begin{multline}\label{eq:eigvals}
        E_{\pm} = \frac{1}{2}(E_1 + E_2 + W_1 + W_2) \\ 
            \pm \frac{1}{2}\left[(E_1 - E_2 + W_1 - W_2)^2 + 4|W|^2 \right]^{1/2} \;,
    \end{multline}
    which are the energies of the confined atom. The matrix elements $W_1$, $W_2$ and $W$ depend on the potential depth $U_0$ as well as the position and width of the cage wall: 
    \begin{eqnarray}\label{eq:matels1}
        W_1 &=& U_0 \int_{\Omega} \int_{r_1}^{r_2} |\phi_{1s}(\mathbf{r})|^2 d^3\mathbf{r} \;, \\
        \label{eq:matels2}
        W_2 &=& U_0 \int_{\Omega} \int_{r_1}^{r_2} |\phi_{2s}(\mathbf{r})|^2 d^3\mathbf{r} \;, \\
        \label{eq:matels3}
        W &=& U_0 \int_{\Omega} \int_{r_1}^{r_2} \phi_{1s}^{*}(\mathbf{r}) \phi_{2s}(\mathbf{r}) d^3\mathbf{r} \; .
    \end{eqnarray}
    Although the radial part of the integrals are over the width of the cage wall, $r\in [r_1,r_2]$, the angular integrals are evaluated over the  entire $4\pi$ solid angle since the potential is spherically symmetric. This decouples states with different parity, such as 2s$^*$ and 2p$^*$, and $W_{2s^*,2p^*}=0$. As a result, the hamiltonian in Eq.~\eqref{eq:hmat} becomes diagonal with eigenvalues $E_1+W_1$ and $E_2+W_2$. 

    From Eq.~\ref{eq:eigvals}, the energies $E_+$ and $E_{-}$ can only become equal when
    \begin{eqnarray}
        E_1 - E_2 + W_1 - W_2 &=& 0 \;, \\
        W &=& 0 \;,
    \end{eqnarray}
    simultaneously. A non-vanishing off-diagonal matrix element prevents the energies from becoming equal for any value of the parameter $U_0$. In this case we observe an avoided crossing. 

    Even though the radial wavefunctions for the 1s$^*$ and 2s$^*$ states are orthogonal, we are only integrating over $r$ within the cage wall. It is clear in Fig.~\ref{fig:diagram} that both states have non-vanishing amplitudes inside the wall. Therefore none of the matrix elements $W_1$, $W_2$ and $W$ vanish. It is also clear that $W_1,W \ll W_2$ since only the exponential tail of the 1s state fall under the potential, whereas $\langle r\rangle =6$ a.u. falls inside the cage wall for the 2s state. These overlaps depend on the position, the width, and the depth of the cage potential. For a given cage radius and width, the matrix elements become: 
    \begin{eqnarray}
        W_1 &=& U_0 K_1 \;, \\
        W_2 &=& U_0 K_2 \;, \\
        W   &=& U_0 K \;, 
    \end{eqnarray}
    where the cage depth $U_0$ is the only a parameter we vary. Here $K_1 \simeq 1.6\times 10^{-3}$, $K_2 \simeq 0.5$, and $K\simeq -2.8\times 10^{-2}$. Note that $K_1,\; |K| \ll K_2$ as expected. The hamiltonian in~\eqref{eq:hmat} now becomes 
    \begin{equation}\label{eq:hmat2}
        \begin{pmatrix}
            E_1+W_1  & W \\
            W^*  & E_2+W_2 
        \end{pmatrix} 
        =
        \begin{pmatrix}
            E_1+ U_0 K_1  & U_0 K \\
            U_0 K  & E_2+ U_0 K_2 
        \end{pmatrix} \;.
    \end{equation}
    We now consider how the eigenvalues of this matrix change as $U_0$ is varied. When $U_0$ is small enough such that the off-diagonal elements are much smaller than the diagonal elements, the matrix is essentially diagonal and $E_1$ and $E_2$ are shifted linearly for small values of $U_0$ (see Fig.~\ref{fig:enrCurves}). Since  $K_1$ is also very small, $E_1+ U_0 K_1$ mainly remains unchanged and $E_2+ U_0 K_2$ decreases as $U_0$ is increased (as the cage deepens, recall that $U_0=-|U_0|$). 

    Around $\sim$0.7 a.u. (point (b) in Fig.~\ref{fig:enrCurves}), the discriminator in Eq.~\eqref{eq:eigvals} is large enough to give two distinct energies, $E_+$ and $E_{-}$. This prevents the energies from becoming equal, and the levels repel. The size of the avoided crossing is $\Delta_{12}=|E_+ - E_{-}|=\sqrt{(E_1 - E_2 + W_1 - W_2)^2 + 4|W|^2 }$. At this point, the eigenvectors switch roles and $K_1$ starts to behave like $K_2$ and $K_2$ like $K_1$. As a result, as $U_0$ is decreased further, 1s$^*$ starts to mimic 2s$^*$ prior to the avoided crossing, and 2s$^*$ starts to look like the former 1s$^*$. This is what we see in Fig.~\ref{fig:enrCurves} from $\sim$0.7 a.u. to $\sim$2.6 a.u., which is reflected in the radial distributions of the states 1s$^*$ and 2s$^*$ in Fig.~\ref{fig:2s-wf}. As these states go through the first avoided crossing in Fig.~\ref{fig:enrCurves}, the 1s$^*$ and 2s$^*$ radial wave functions start exchanging their spatial profiles (Fig.~\ref{fig:2s-wf}(c)). The radial states stay like these until the 2s$^*$ state meets the 3s$^*$ state at the next avoided crossing around 2.6 a.u. in Fig.~\ref{fig:enrCurves}. 

    The level crossings described above are responsible for three distinctly different regimes separated by the two avoided crossings seen in Fig.~\ref{fig:enrCurves}: (1) before the first avoided crossing at $\sim$0.7 a.u., (2) between the two avoided crossings, and (3) after the second avoided crossing at $\sim$2.6 a.u.. In simulations of HHG out of H(2s$^*$)@C$_{60}$, the depth of the fullerene cage is 0.3 a.u., which is labeled (a) in Fig.~\ref{fig:enrCurves}. To demonstrate how the HHG spectra change as we vary the cage depth across the avoided crossing, we have performed simulations at two other cage depths at 0.7 a.u. and 1.3 a.u. labeled as (b) and (c) in Fig.~\ref{fig:enrCurves}. The radial distributions of the 1s$^*$ and the 2s$^*$ states at these three points are shown in Fig.~\ref{fig:2s-wf}. Before the avoided crossing at (a), the radial distribution of the 2s$^*$ state is mainly localized on the fullerene cage wall and the 1s$^*$ state is entirely inside. For a laser intensity of $5\times 10^{13}$/$n^8$ W/cm$^2$ and wavelength $800 n^3$ nm in these set of simulations, the ground state generates no high-harmonics.

    On the other hand, the 2s$^*$ state generates very few harmonics with a much smaller cut-off frequency compared to a free atom in the 2s state (Fig.~\ref{fig:hhg-h2s}(a)). This is, again, due to that fact that the 2s$^*$ state can tunnel to either to the left or the right of the fullerene cage wall due to its radial localization. This severely reduces the amplitude of the tunneled wave packet that can effectively contribute to the HHG process. During the avoided crossing at (b), the 1s$^*$ and the 2s$^*$ states exchange their radial localization, while retaining their nodal structures. After the avoided crossing at point (c), the 2s$^*$ state is localized essentially where the 1s$^*$ state used to be. In this case, the binding energy of the initial state and the width of the tunneling barrier are too large for the 2s$^*$ state, and HHG is turned off completely. There is no effective HHG from the 2s$^*$ state until after the second avoided crossing with the 3s$^*$ state at $\sim$2.6 a.u.. Following the second avoided crossing, the 2s$^*$ state becomes localized on the cage again, and the harmonic spectra reverts to a spectrum similar to the one in Fig.~\ref{fig:hhg-h2s}(a). The overall intensity of the generated harmonics decrease, however, since the 2s$^*$ state becomes more deeply bound on the cage wall, and tunneling becomes progressively more difficult. 
    
     \begin{figure}[h!tb]
            \begin{center}
                \resizebox{0.99\columnwidth}{!}{\includegraphics[angle=0]{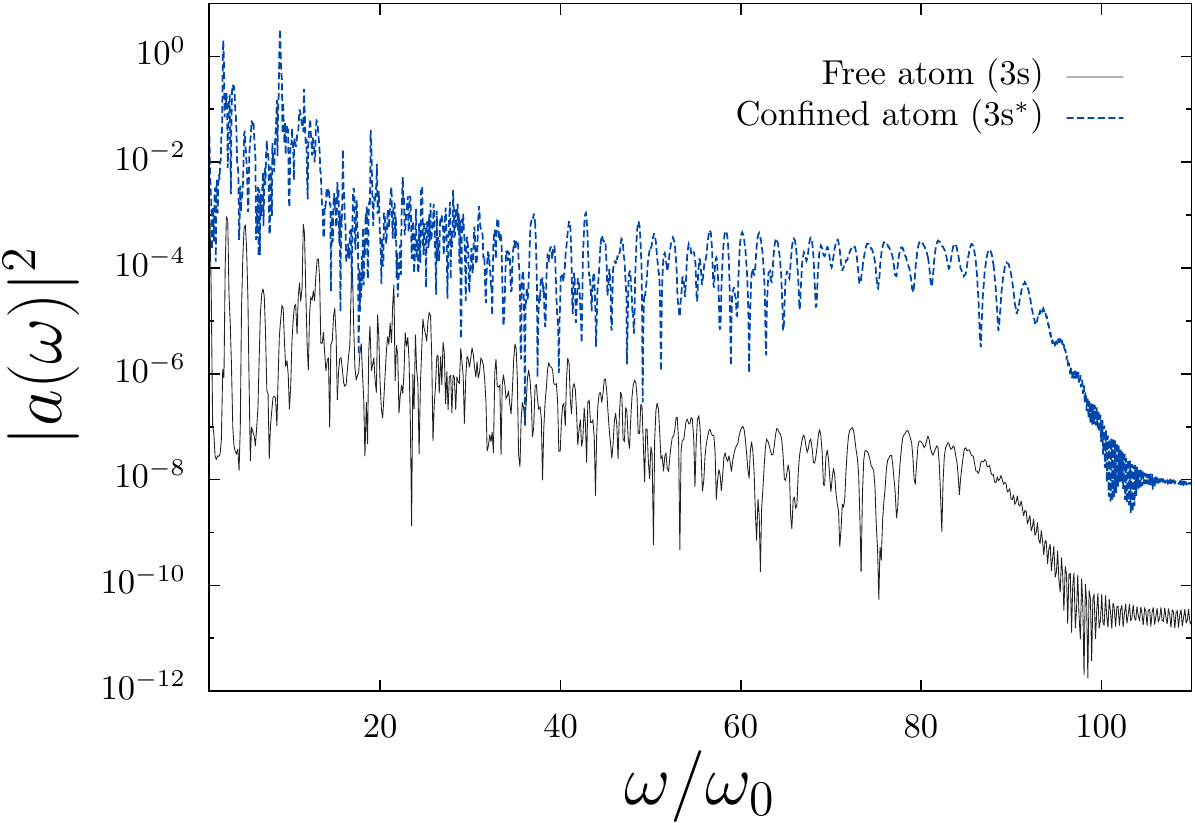}}
            \end{center}
            \caption{(Color online) Harmonic spectra from free H (black solid) and H@C$_{60}$ (blue dashed) when they are initially prepared in the 3s and 3s$^*$ states. The intensity and wavelength are $5.0\times 10^{13}$/$n^8$ W/cm$^2$ and $800 n^3$ nm where $n=3$. The enhancement results from confining the H atom inside the C$_{60}$ shell, despite the fact that the 3s$^*$ state is more deeply bound compared to the 3s state. There is $\sim$4\% ionization from the confined system whereas there is virtually no ionization from the free atom. 
            }
            \label{fig:hhg-h3s}
    \end{figure}

	\begin{figure*}[h!tb]
            \begin{center}
                \resizebox{1.0\textwidth}{!}{\includegraphics[angle=0]{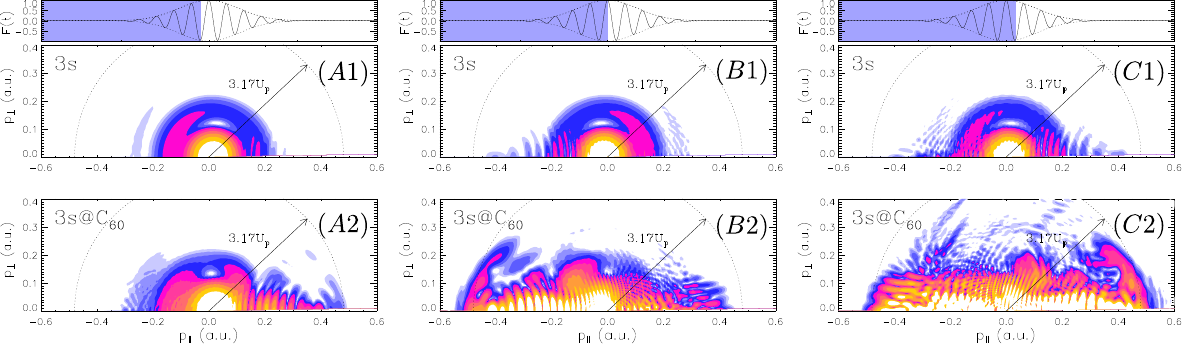}}
            \end{center}
            \caption{(Color online) Time-dependent momentum distribution of the total wavefunction evaluated according to the method described in Sec.~\ref{sec:theory:pmap}. A1, B1 and C1 are for a free H atom prepared in the 3s state at three specific points spanning one laser period at the peak of the laser pulse (indicated on top). A2, B2 and C2 show the momentum distributions at the same instances during the pulse for H(3s$^*$)@C$_{60}$. The boundary which marks the kinetic energy corresponding to the cut-off frequency for a free atom at 3.17$U_p$ is indicated as a dashed semicircle in each case. 
            }
            \label{fig:pmap}
    \end{figure*}
    
	\subsection{H(3s$^*$)@C$_{60}$}\label{subsec:h-3s}
	An intriguing situation arises when we compare spectra from systems initially prepared in excited states whose classical turning points at $2n^2$ as well as $\langle r\rangle$ lie outside the fullerene shell. In this case, we see substantial enhancement in the spectrum from the atom-fullerene system compared with the spectrum from a free H atom. Specifically, we report results for the 3s and 3s$^*$ states here where the intensity and the wavelength are $5.0\times 10^{13}$/$n^8$ W/cm$^2$ and $800 n^3$ nm with $n=3$. In the 3s$^*$ state, the classical turning point is at $\sim$18 a.u. and $\langle r\rangle$$\sim$13.5 a.u., both of which are well outside the confining C$_{60}$ shell. We see $\sim$4 orders of magnitude enhancement in the HHG yield from H(3s$^*$)@C$_{60}$ compared to the free H atom alone (Fig.~\ref{fig:hhg-h3s}). We see similar behavior for higher excited states as well, such as when we compare 4s and 4s$^*$ states, which we do not report here. We should also note here that the enhancement seen in Fig.~\ref{fig:hhg-h3s} does not disappear when we increase the intensity as was the case for HHG yield from the 1s$^*$ state in Fig.~\ref{fig:hhg-h1s}(b). 

	Comparing dipole $|d(\omega)|^2$, velocity $|v(\omega)|^2$ and acceleration $|a(\omega)|^2$ forms of the HHG spectra  from the free H atom, we find good agreement between all three forms until the cut-off seen in Fig.~\ref{fig:hhg-h3s}. In the confined system, all three forms agree very well in the low harmonics region where $\omega/\omega_0 \lesssim 15$ after which both the dipole and the velocity forms start to differ from the acceleration form. Particularly, the dipole form  becomes substantially different from the other two beyond $\omega/\omega_0$$\sim$15. The large difference in $\omega^4 |\langle z\rangle(\omega)|^2$ has to do with substantial oscillations in $\langle z\rangle(t)$ following the laser pulse resulting from significant mixing in the final state. Using the Green's function method from Sec.~\ref{sec:theory:greens}, we determine how much of the amplitude participating in the HHG process is mixed into adjacent states after the laser pulse. We find that the laser pulse leaves the atom in a superposition state involving $n=4$ states in addition to states in the $n=3$ manifold. In fact, the final state of the confined H atom includes comparable amplitudes from both $n$-manifolds. In contrast, we see no mixing in the final state when the H atom is free. As a result, $\langle z\rangle(t)$ exhibits substantial oscillations In the final state of the system. We see this behavior neither in the free H atom nor in the 1s$^*$ and 2s$^*$ states of H@C$_{60}$, in which case all three forms of the spectra agree well until the cut-off harmonics. 
	
	The difference between the three forms of the harmonic spectra beyond $\omega/\omega_0$$\sim$15 is muted when we compare them after arranging the pulse length such that $\langle z\rangle(t)$, $\langle \dot{z}\rangle(t)$ and $\langle \ddot{z}\rangle(t)$ vanish at the end of the pulse in their {\it respective} calculations. Particularly, the velocity and acceleration forms agree well in this case, whereas the dipole form still differs beyond $\omega/\omega_0$$\sim$15. 
	
	We want to emphasize that the enhancement in Fig.~\ref{fig:hhg-h3s} takes place despite two facts: (1) The 3s$^*$ state is more deeply bound than the 3s state, and their energies differ by $\sim$3.5\%. Despite tunneling out of a more deeply bound state, the H(3s$^*$)@C$_{60}$ system emits a far more intense plateau of harmonic photons. (2) There is virtually no ionization out of the 3s state whereas the ionization from H(3s$^*$)@C$_{60}$ is at the $\sim$4\% level. This also works towards reducing the HHG yield from the H(3s$^*$)@C$_{60}$ relative to the yield from a free H atom. 
	
	To understand the origin of the enhancement seen in Fig.~\ref{fig:hhg-h3s}, first note that the enhancement throughout the plateau is essentially uniform, meaning that the enhanced emission yield at high-order harmonics has not come at the expense of the lower harmonics: there is enhancement throughout the plateau between $\sim$2-4 orders of magnitude. This suggests that the significant enhancement seen in Fig.~\ref{fig:hhg-h3s} cannot result from a higher-order process where low order harmonics become converted to higher harmonics. Furthermore, the existence of the fullerene shell cannot alter the dynamics of the tunneled wave-packet in the propagation step of the HHG process, which happens well outside the C$_{60}$ cage. Combined with the fact that the H(3s$^*$)@C$_{60}$ system effectively suffers from a smaller tunneling rate, the enhancement in Fig.~\ref{fig:hhg-h3s} could only have come from the recombination step of the HHG process. 

	\begin{figure}
		\begin{center}
			\resizebox{1.0\columnwidth}{!}{\includegraphics[angle=0]{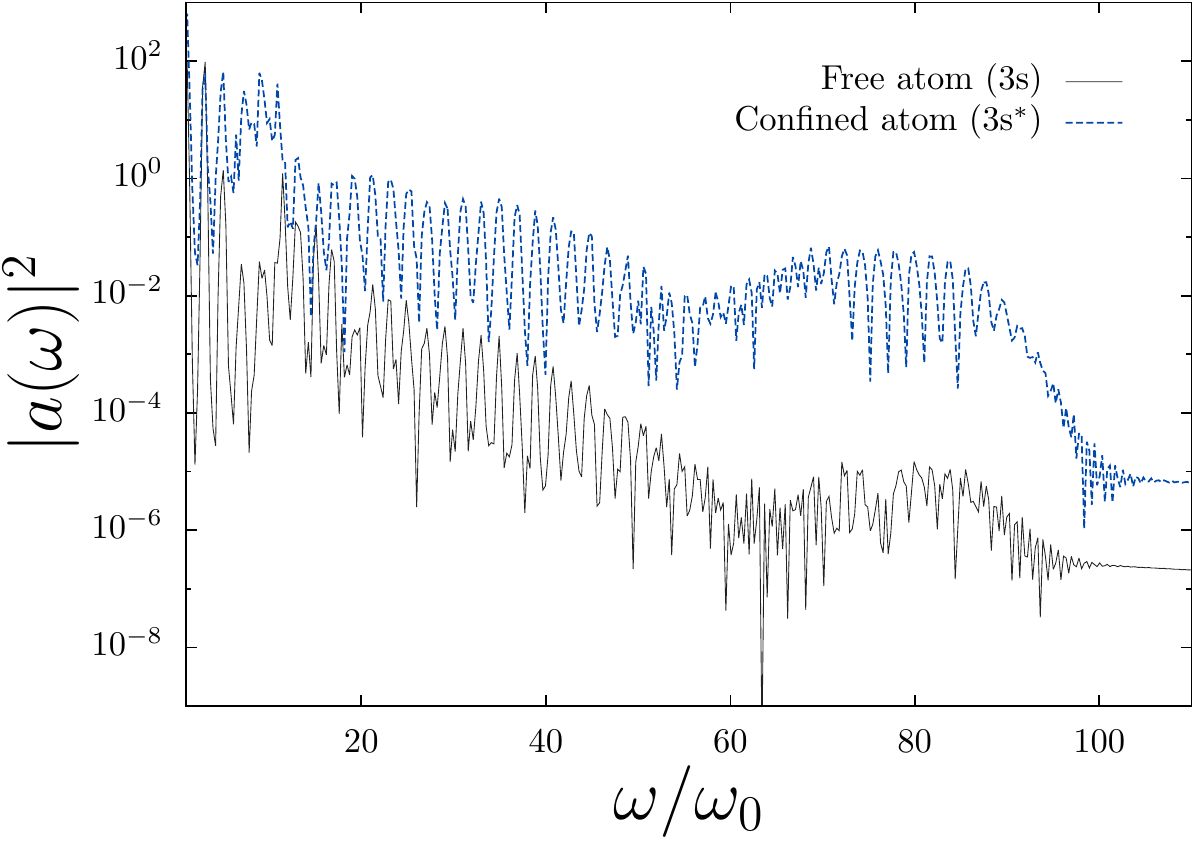}}
		\end{center}
		\caption{(Color online) Spectra reported in Fig.~\ref{fig:hhg-h3s} except calculated using the Green's function method described in Sec.~\ref{sec:theory:greens} for the same laser parameters. 
		}
		\label{fig:hhg-h3s-gfunc}
	\end{figure}
	
   Comparing the momentum distributions for the free H and H(3s$^*$)@C$_{60}$, we can gain insight into the key differences in the dynamics taking place in both cases to help elucidate the mechanism at play leading to the enhancement. We evaluate momentum distributions at three instances during the laser pulse, which are shown in Fig.~\ref{fig:pmap}. We take these snapshots over a complete cycle of the laser field at the peak of the pulse envelope (shown on top). Panels A1, B1 and C1 are for the free H atom, and the panels A2, B2 and C2 are for H(3s$^*$)@C$_{60}$. The first thing we notice is that in both cases the momentum is confined within a sphere in the momentum space with a radius corresponding to a kinetic energy of 3.17$U_p$ (dashed circle). This aligns with the fact that both HHG spectra seen in Fig.~\ref{fig:hhg-h3s} have the same cut-off frequency and no higher harmonics have been generated by confining the atom inside the fullerene. The second point is that there is a larger momentum spread in the $p_{\perp}$ component for H(3s$^*$)@C$_{60}$ compared to the free atom momentum distribution, which is important in understanding the origin of the enhancement seen in Fig.~\ref{fig:hhg-h3s}. The larger the momentum spread in the direction perpendicular to the laser polarization the smaller the spatial spread of the wavepacket in this direction, which would lead to more efficient recombination. 
   
   \begin{figure*}
		\begin{center}$
			\begin{array}{ccc}  
				  \resizebox{0.3\textwidth}{!}{\includegraphics[angle=0]{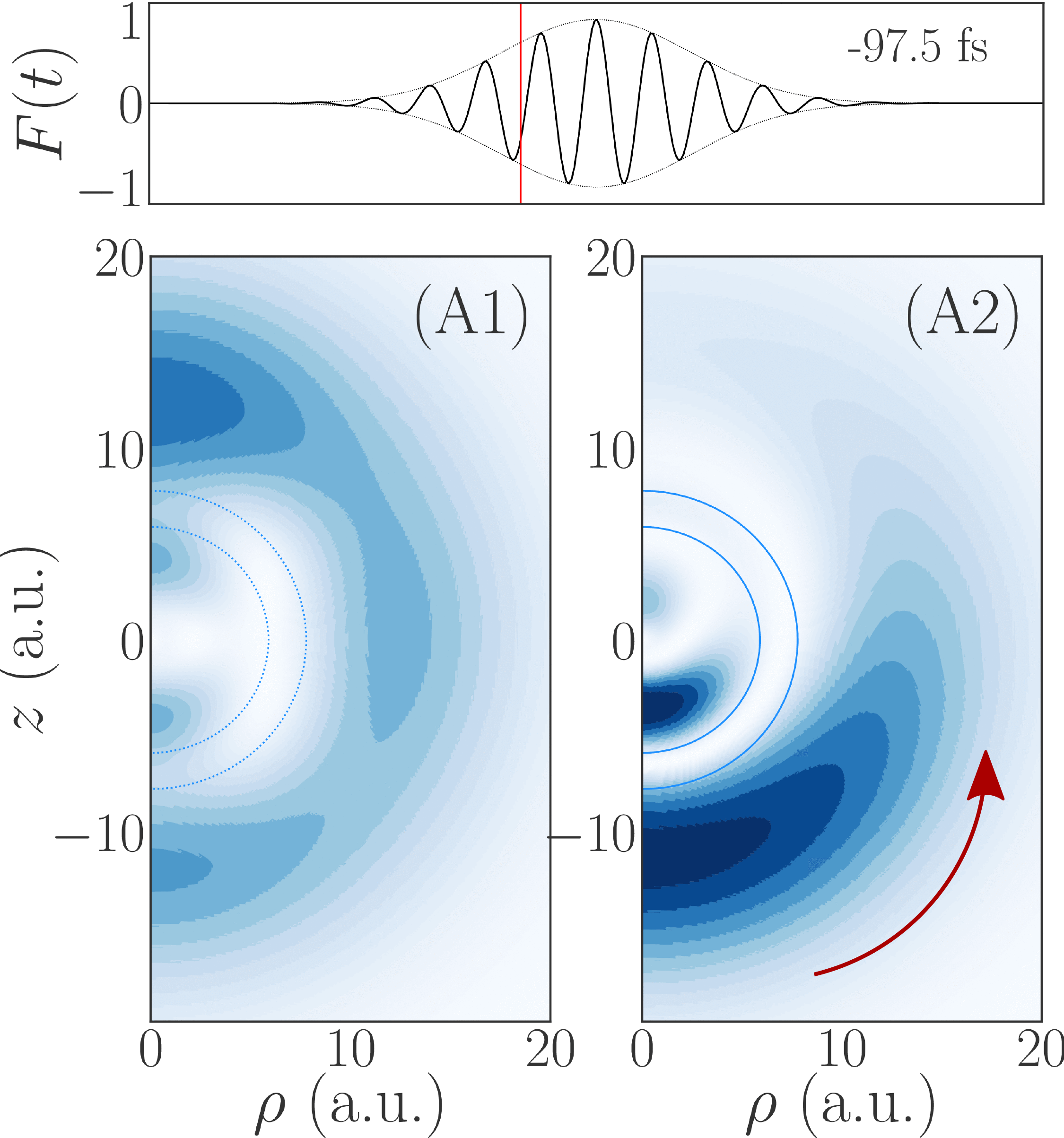}} &
				  \resizebox{0.3\textwidth}{!}{\includegraphics[angle=0]{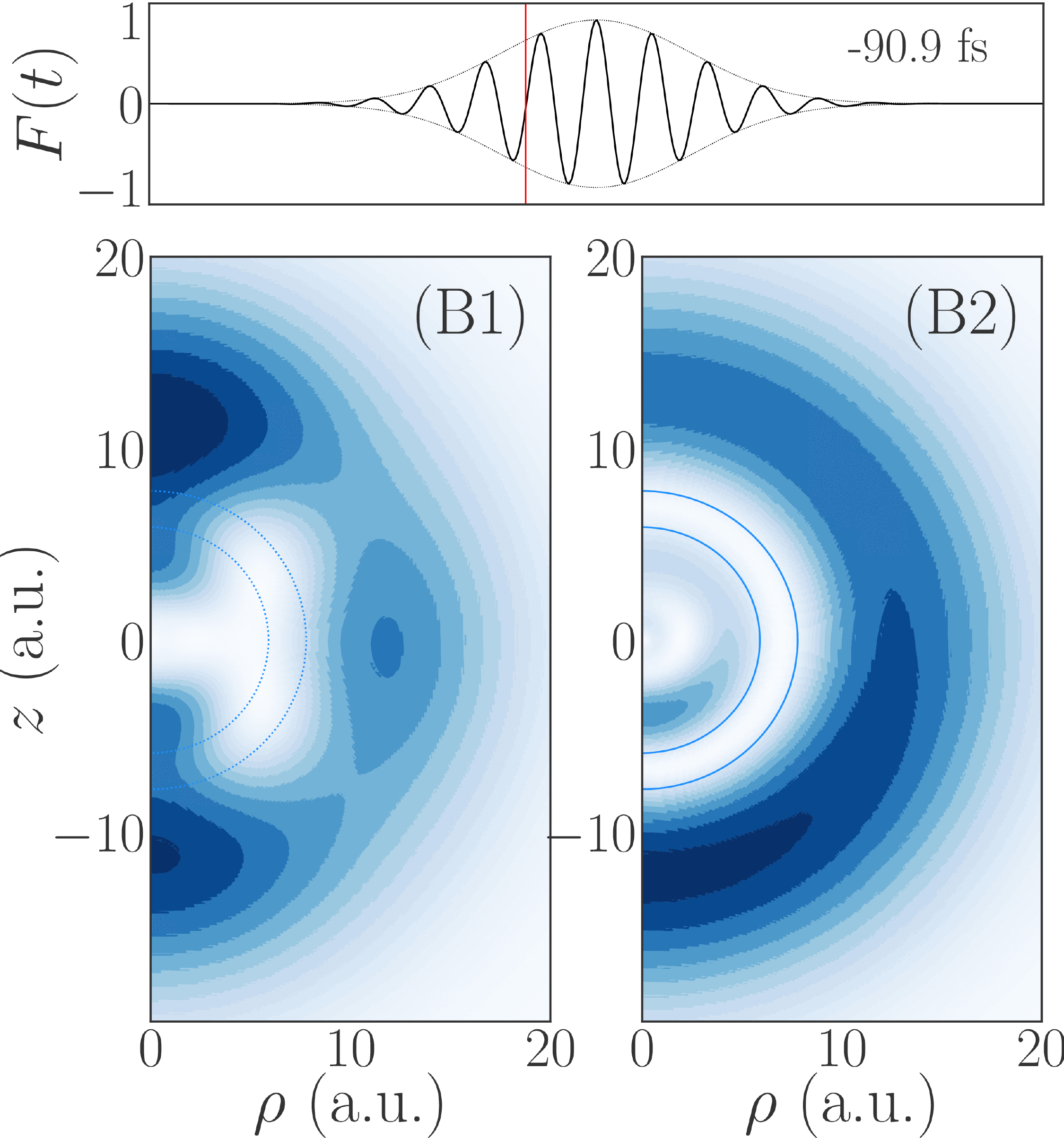}} &
				  \resizebox{0.3\textwidth}{!}{\includegraphics[angle=0]{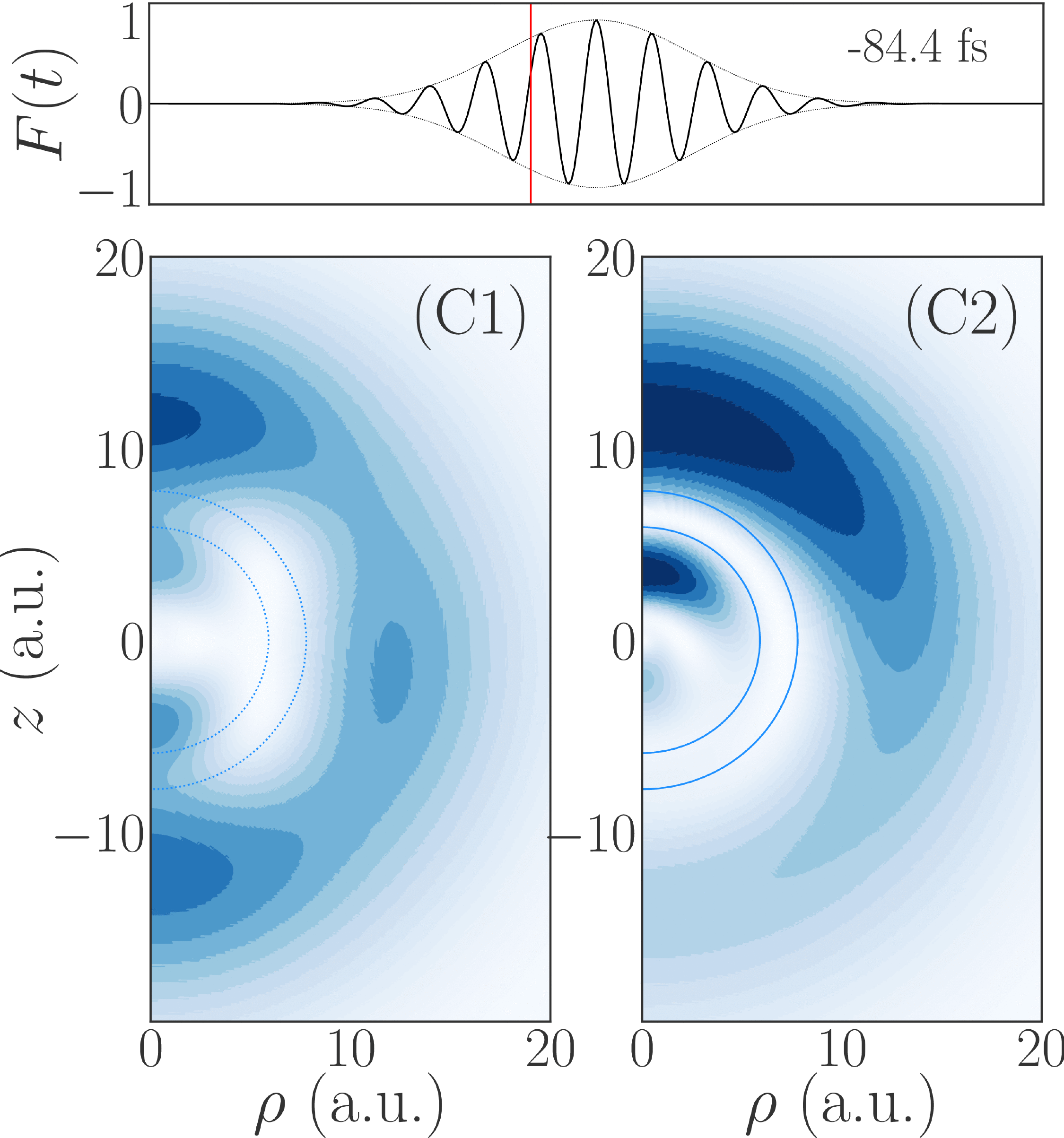}} 
			\end{array}$
		\end{center}
		\caption{(Color online) Wavefunction plots showing the squeezing of the returning wave packet before recombination for the $n=3$ initial state. Each snapshot contains two panels. A1, B1 and C1 show the PCD from H(3s), and A2, B2 and C2 show PCD from the H(3s$^*$)@C$_{60}$ system. The position of the fullerene shell is marked by dashed (free atom) and solid (confined) blue semicircles in each panel. The snapshots are taken over a complete laser period instances of which we indicate on top of each time-frame. 
		}
		\label{fig:wfunc}
	\end{figure*}

	When the classical turning point of the initial state is outside the fullerene shell, such as for states with $n\ge 3$, tunneling happens outside the C$_{60}$ cage ($2n^2 > r_c$), but recombination still takes place inside (momentum conservation). This means, the outgoing wavepacket is the same as that from a free atom, but when it returns to its parent ion to recombine, it must go inside the shell. When it does, the wavepacket is spatially squeezed in the direction perpendicular to the laser polarization by the potential shell. The momentum distribution doesn't exhibit higher momentum components compared to that for a free atom (Fig.~\ref{fig:pmap}), and the incoming wavepacket having a smaller spatial spread means a higher probability for recombination. This is why we do not see any higher harmonics generated in the confined case; the cut-off remains exactly the same from when the atom was free. This is evident from Fig.~\ref{fig:pmap} where the momentum distribution remains within a sphere with a radius corresponding to the maximum kinetic energy of 3.17$U_p$ in the momentum space in both cases. However, the intensity of the emitted harmonics is increased, therefore, the recombination efficiency must have been enhanced. This is also implied by the fact that no harmonics lose intensity compared with the free atom in Fig.~\ref{fig:hhg-h3s}, where intensity is larger for all the harmonics (or at least the same). This suggests that the effect cannot be the result of some of the harmonics being converted into others. See Fig.~\ref{fig:3cases} for all three distinct cases discussed. 

    It should be possible to see the lensing effect depicted in Fig.~\ref{fig:3cases} leading to the enhancement in Fig.~\ref{fig:hhg-h3s} by comparing the time-dependent probability current densities (PCD) $|\psi(\mathbf{r},t)|^2$ for H(1s) and H(3s$^*$)@C$_{60}$ in space. The problem, however, is that only a small fraction of the total wavefunction contributes to the HHG process, and most of the amplitude sits where the initial state resides. The fullerene shell also lies in this region of space, and any interesting dynamics taking place here is buried underneath the initial amplitude, and therefore, is not visible. One way around this difficulty is to use the Green's function method described in Sec.~\ref{sec:theory:greens}. This is essentially the time-dependent perturbation theory in which we solve for the time-evolution of the first-order correction to the total wavefunction, $\psi_1(\mathbf{r},t)$ from Sec.\ref{sec:theory:greens}. This part of the wavefunction is what mainly contributes to the HHG process. Although no net photon absorption/emission takes place in the tunneling step of HHG, time-dependent perturbation theory can still describe tunneling if a time-dependent potential~\cite{Rei95} enables it. The time-dependent perturbation theory is routinely used in calculating transmission/reflection coefficients at solid state junctions, and it gives accurate results. We also calculate $|a(\omega)|^2$ using $\psi_1(\mathbf{r},t)$ to ensure that the Green's function method is capable of capturing the relevant physics, and can reproduce the enhancement we see in Fig.~\ref{fig:hhg-h3s}. 

    The HHG spectra from H(1s) and H(3s$^*$)@C$_{60}$ obtained using  $\psi_1(\mathbf{r},t)$ are compared in Fig.~\ref{fig:hhg-h3s-gfunc}. The perturbative method we use qualitatively reproduces the enhancement effect we see in Fig.~\ref{fig:hhg-h3s}.  There is $\sim$4 orders of magnitude enhancement for harmonics above $\omega/\omega_0 > 60$, and between 1-4 orders of magnitude enhancement for lower harmonics. For no harmonic in the plateau region, there is a reduction in the emitted photon intensity from H(3s$^*$)@C$_{60}$ compared to the free atom. This shows that the physical mechanism responsible for the enhancement we see in Fig.~\ref{fig:hhg-h3s} is, in essence, captured by the Green's function method. 

    We plot the time-dependent PCD at four instances during the laser pulse in Fig.~\ref{fig:wfunc}. The four snapshots span one laser cycle in the rising edge of the pulse as indicated on top for each time-frame. The left panels (A1, B1 and C1) show the PCD for a free atom, and right panels (A2, B2 and C2) show the same for the H@C$_{60}$ system. The position of the spherically symmetric fullerene shell is marked by the dashed and solid circles in each case. Note that the PCDs are not synchronized in the snapshots because the binding energies for the 3s and 3s$^*$ states are slightly different, which translates into different tunneling times. Fig.~\ref{fig:wfunc} shows that when the returning wavepacket arrives at the fullerene cage in the first frame, the transmitted part of the wavefunction is denser inside the cage (A2) relative to the same event for the free atom (A1). This part of the wavepacket is revived on the opposite side of the atomic core inside the cage in the second half of the laser cycle (C1 and C2). At no point during the laser cycle, the transmitted part of the wavepacket for the free atom is as densely focused along the laser polarization as for the atom confined inside the C$_{60}$ shell. This is in agreement with the time-dependent momentum distributions we see in Fig.~\ref{fig:pmap} which show a much broader spread of momentum in the direction perpendicular to the laser polarization for H(3s$^*$)@C$_{60}$. A larger spread in momentum translates to a narrower spread in space, and the wavepacket shows a stronger peak inside the C$_{60}$ shell. 
    
    \begin{figure}
		\begin{center}
			\resizebox{1.0\columnwidth}{!}{\includegraphics[angle=0]{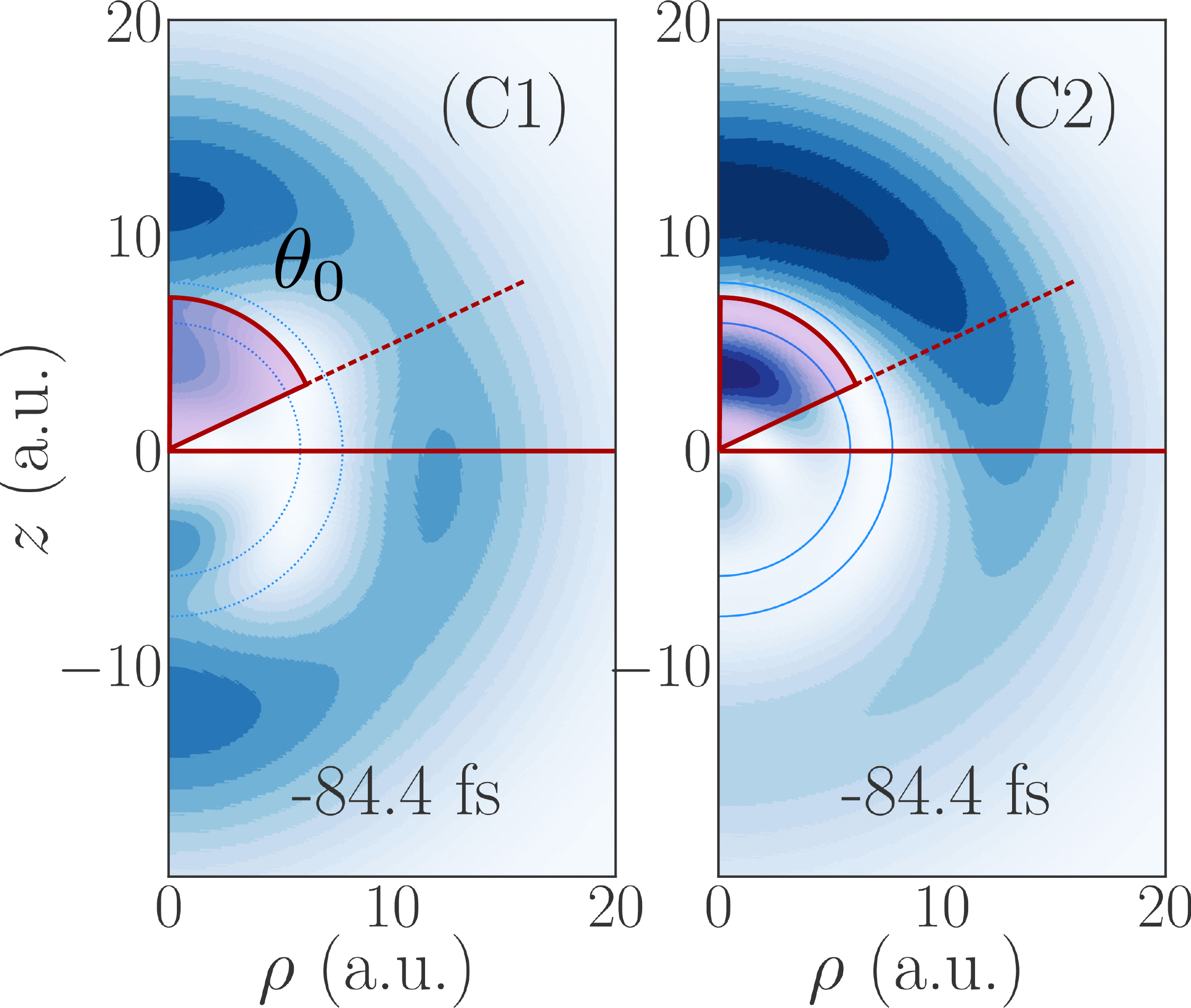}}
		\end{center}
		\caption{(Color online) Figure highlights parts of the PCD near the nucleus obtained using the Green's function method $\sim$85 fs before the peak of the laser pulse for H(3s) and H(3s$^*$)@C$_{60}$ (also seen in Fig.~\ref{fig:wfunc} C1 and C2). The figure overlays a purple region to highlight amplitude near the nucleus, both of which subtended by some the same angle $\theta_0$ in panels C1 and C2. It is clear that there is more amplitude inside the highlighted region in C2 than in C1 -- suggesting a higher probability of recombination. 
		}
		\label{fig:hhg-h3s-gfunc-160}
	\end{figure}

	To clarify this point further, we can look at region inside the C$_{60}$ shell more closely. Fig.~\ref{fig:hhg-h3s-gfunc-160} provides a magnified view of the transmitted parts of the wavefunction depicted in the last frame in Fig.~\ref{fig:wfunc} (C1 and C2).  The purple color highlights the part of the wave packet inside the cage in  Fig.~\ref{fig:hhg-h3s-gfunc-160}(C2). The same region is also highlighted in Fig.~\ref{fig:hhg-h3s-gfunc-160}(C1) where the cage is absent (free atom). The highlighted regions of space in C1 and C2 contain wavepackets approaching their parent ions for rescattering. These rescattering wavepackets partially recombine to emit photons. It is clear that in panel C2, the  highlighted region contains more amplitude than the same region in C1. More amplitude means larger recombination probability, therefore larger intensity of emitted harmonics. Please note that although C1 and C2 are snapshots taken at exactly at the same time, the wavepackets are not exactly synchronized. This is because the binding energies of the 3s and 3s* states are slightly different, which results in different tunneling times. Because of this, rescattering takes place at different times. Closer inspection of the time-evolution shows that this does not change the qualitative picture: the amplitude seen the highlighted region in panel C1 never becomes larger than the amplitude in the same region in C2 throughout the entire time evolution of these systems.

	\section{Future Directions}\label{sec:conclusion}
	In this paper, we have demonstrated that coupling atoms to nanostructures can significantly enhance HHG yield, and we used endofullerenes as an example. However, we used H(3s$^*$)@C$_{60}$ in our demonstration chiefly because of the lower intensities needed to extract high-harmonics from the hydrogen atom, while keeping depletion of atoms due to ionization at a minimum.  Furthermore, the intensities we use help retain structural integrity of the fullerene (C$_{60}$) in which the H atom is confined, because C$_{60}$ easily fragments through a multi-photon ionization process in intensities typical for HHG using noble gas atoms in experiments. The question is then how the enhancement we describe in Sec.\ref{subsec:h-3s} can be made into a useful tool in the laboratory. 
    
	The issue of fragmentation of the fullerene cages is partly alleviated by realizing that even if a significant fraction of C$_{60}$ is destroyed in intensities typical for noble gas atoms well before the peak of the laser pulse, survival of even a small fraction of the initial endofullerene population would mean a significant enhancement in the overall macroscopic response of the medium. This is due to the fact there are $\sim$4 orders of magnitude enhancement in the response from the confined species in Fig.~\ref{fig:hhg-h3s}. Even if $\sim$1\% of Ne@C$_{60}$ or Ar@C$_{60}$ populations survive the peak of the laser pulse, there can still be a substantial increase in the macroscopic response if the medium initially starts with endofullerenes. 
	
	To assess what fraction of the initial endofullerenes needs to survive the laser pulse for a meaningful increase in the yield, one needs to propagate the generated harmonics through the mixed medium consisting of atoms and endofullerenes to obtain the final macroscopic response. Using experimental data available, such as those reported in Ref.~\cite{HunStaHui96}, one can estimate what laser intensities are optimal that can maximize the high-harmonic photon yield while keeping fragmentation rate to a minimum to obtain a meaningful enhancement in the macroscopic spectrum. One issue that needs attention is that ionization from fullerenes will alter the dispersion characteristics of the macroscopic medium. Therefore it is important to take modified refraction index of the medium into account due to the hot carbon plasma generated from fragmentation of the fullerenes. This means it is also necessary to propagate the fundamental laser pulse in parallel with the generated harmonics. It would be interesting to understand how phase-matching works in this case, where the endofullerene gas is gradually transformed to an environment consisting of atoms inside an ionized carbon plasma dispersing the harmonic photons. 
	
	There are other demonstrated mechanisms which can potentially be used to enhance the HHG yield. One example of this is has been demonstrated using plasmonic effects~\cite{plasmonHusakou2011, plasmonKim2008, inhomogYavuz2012, inhomogCiappina2012}. Because the physics is different in both cases, these can be combined to obtain enhancement from two physically independent processes. Another example is preparing the initial atom in a superposition state involving the ground and an excited state~\cite{ZhZhCh11,ZhChYa10}. Coupling the ground state to a Ry state also increases HHG efficiency, partly because the Ry electron can tunnel more efficiently than the ground state, and the returning wavepacket preferentially recombines back into the state out of which it tunneled. 
    
    As one needs intense and high-frequency harmonics for applications, such as attosecond pulse generation, ground state atoms are usually preferred in HHG rather than the excited states. Starting with a superposition state of the combined atom-fullerene system can result in enhanced harmonic photon yield compared to using only the ground state endofullerenes. Furthermore, it may be interesting to see how all these different enhancement mechanisms, {\it e.g.} plasmonic effects, superposing with a Ry state, and confinement can compound to give an enhanced yield in the macroscopic response of a gaseous target.

	\section{Acknowledgement}
	TT was supported in part by the National Science Foundation (NSF) Grant No. PHY-1212482 and acknowledges the use of UNR Grid Cluster for simulations. EAB was supported by BAPKO of Marmara University Grant No. FEN-A-071015-0477. EAB also thanks Mr. Dogan Bolak for providing additional computational resources.

	\bibliography{erdi-hhg,ttopcu-hhg}

\begin{thebibliography}{29}
\expandafter\ifx\csname natexlab\endcsname\relax\def\natexlab#1{#1}\fi
\expandafter\ifx\csname bibnamefont\endcsname\relax
  \def\bibnamefont#1{#1}\fi
\expandafter\ifx\csname bibfnamefont\endcsname\relax
  \def\bibfnamefont#1{#1}\fi
\expandafter\ifx\csname citenamefont\endcsname\relax
  \def\citenamefont#1{#1}\fi
\expandafter\ifx\csname url\endcsname\relax
  \def\url#1{\texttt{#1}}\fi
\expandafter\ifx\csname urlprefix\endcsname\relax\def\urlprefix{URL }\fi
\providecommand{\bibinfo}[2]{#2}
\providecommand{\eprint}[2][]{\url{#2}}

\bibitem[{\citenamefont{Lein}(2007)}]{imaging1Lein2007}
\bibinfo{author}{\bibfnamefont{M.}~\bibnamefont{Lein}},
  \bibinfo{journal}{Journal of Physics B: Atomic, Molecular and Optical
  Physics} \textbf{\bibinfo{volume}{40}}, \bibinfo{pages}{R135}
  (\bibinfo{year}{2007}), ISSN \bibinfo{issn}{0953-4075}.

\bibitem[{\citenamefont{Hentschel et~al.}(2001)\citenamefont{Hentschel,
  Kienberger, Spielmann, Reider, Milosevic, Brabec, Corkum, Heinzmann,
  Drescher, and Krausz}}]{imaging2Hentschel2001}
\bibinfo{author}{\bibfnamefont{M.}~\bibnamefont{Hentschel}},
  \bibinfo{author}{\bibfnamefont{R.}~\bibnamefont{Kienberger}},
  \bibinfo{author}{\bibfnamefont{C.}~\bibnamefont{Spielmann}},
  \bibinfo{author}{\bibfnamefont{G.~a.} \bibnamefont{Reider}},
  \bibinfo{author}{\bibfnamefont{N.}~\bibnamefont{Milosevic}},
  \bibinfo{author}{\bibfnamefont{T.}~\bibnamefont{Brabec}},
  \bibinfo{author}{\bibfnamefont{P.}~\bibnamefont{Corkum}},
  \bibinfo{author}{\bibfnamefont{U.}~\bibnamefont{Heinzmann}},
  \bibinfo{author}{\bibfnamefont{M.}~\bibnamefont{Drescher}}, \bibnamefont{and}
  \bibinfo{author}{\bibfnamefont{F.}~\bibnamefont{Krausz}},
  \bibinfo{journal}{Nature} \textbf{\bibinfo{volume}{414}},
  \bibinfo{pages}{509} (\bibinfo{year}{2001}), ISSN \bibinfo{issn}{00280836}.

\bibitem[{\citenamefont{Yang et~al.}(2015)\citenamefont{Yang, Li, Zhang, and
  Lin}}]{Yang2015}
\bibinfo{author}{\bibfnamefont{Y.-Y.} \bibnamefont{Yang}},
  \bibinfo{author}{\bibfnamefont{Q.-G.} \bibnamefont{Li}},
  \bibinfo{author}{\bibfnamefont{L.}~\bibnamefont{Zhang}}, \bibnamefont{and}
  \bibinfo{author}{\bibfnamefont{X.-C.} \bibnamefont{Lin}},
  \bibinfo{journal}{Plasmonics}  (\bibinfo{year}{2015}), ISSN
  \bibinfo{issn}{1557-1955},
  \urlprefix\url{http://link.springer.com/10.1007/s11468-015-0010-7}.

\bibitem[{\citenamefont{McGrath et~al.}(2014)\citenamefont{McGrath, Hawkins,
  Simpson, Siegel, Diveki, Austin, Zair, Castillejo, and
  Marangos}}]{procMcGrath2014}
\bibinfo{author}{\bibfnamefont{F.}~\bibnamefont{McGrath}},
  \bibinfo{author}{\bibfnamefont{P.}~\bibnamefont{Hawkins}},
  \bibinfo{author}{\bibfnamefont{E.}~\bibnamefont{Simpson}},
  \bibinfo{author}{\bibfnamefont{T.}~\bibnamefont{Siegel}},
  \bibinfo{author}{\bibfnamefont{Z.}~\bibnamefont{Diveki}},
  \bibinfo{author}{\bibfnamefont{D.}~\bibnamefont{Austin}},
  \bibinfo{author}{\bibfnamefont{A.}~\bibnamefont{Zair}},
  \bibinfo{author}{\bibfnamefont{M.}~\bibnamefont{Castillejo}},
  \bibnamefont{and} \bibinfo{author}{\bibfnamefont{J.~P.}
  \bibnamefont{Marangos}}, \bibinfo{journal}{Proc. SPIE 8984, Ultrafast
  Phenomena and Nanophotonics XVIII} \textbf{\bibinfo{volume}{89841B}}
  (\bibinfo{year}{2014}).

\bibitem[{\citenamefont{Pabst and Santra}(2013)}]{PaSa13}
\bibinfo{author}{\bibfnamefont{S.}~\bibnamefont{Pabst}} \bibnamefont{and}
  \bibinfo{author}{\bibfnamefont{R.}~\bibnamefont{Santra}},
  \bibinfo{journal}{Phys. Rev. Lett.} \textbf{\bibinfo{volume}{111}},
  \bibinfo{pages}{233005} (\bibinfo{year}{2013}),
  \urlprefix\url{https://link.aps.org/doi/10.1103/PhysRevLett.111.233005}.

\bibitem[{\citenamefont{Shiner et~al.}(2011)\citenamefont{Shiner, Schmidt,
  Trallero-Herrero, Wörner, Patchkovskii, Corkum, Kieffer, Légaré, and
  Villeneuve}}]{ShScTr11}
\bibinfo{author}{\bibfnamefont{A.~D.} \bibnamefont{Shiner}},
  \bibinfo{author}{\bibfnamefont{B.~E.} \bibnamefont{Schmidt}},
  \bibinfo{author}{\bibfnamefont{C.}~\bibnamefont{Trallero-Herrero}},
  \bibinfo{author}{\bibfnamefont{H.~J.} \bibnamefont{Wörner}},
  \bibinfo{author}{\bibfnamefont{S.}~\bibnamefont{Patchkovskii}},
  \bibinfo{author}{\bibfnamefont{P.~B.} \bibnamefont{Corkum}},
  \bibinfo{author}{\bibfnamefont{J.-C.} \bibnamefont{Kieffer}},
  \bibinfo{author}{\bibfnamefont{F.}~\bibnamefont{Légaré}}, \bibnamefont{and}
  \bibinfo{author}{\bibfnamefont{D.~M.} \bibnamefont{Villeneuve}},
  \bibinfo{journal}{Nature Physics} \textbf{\bibinfo{volume}{7}},
  \bibinfo{pages}{464–467} (\bibinfo{year}{2011}),
  \urlprefix\url{http://dx.doi.org/10.1038/nphys1940}.

\bibitem[{\citenamefont{Gaarde et~al.}(1999)\citenamefont{Gaarde, Salin,
  Constant, Balcou, Schafer, Kulander, and L'Huillier}}]{GaSaCo99}
\bibinfo{author}{\bibfnamefont{M.~B.} \bibnamefont{Gaarde}},
  \bibinfo{author}{\bibfnamefont{F.}~\bibnamefont{Salin}},
  \bibinfo{author}{\bibfnamefont{E.}~\bibnamefont{Constant}},
  \bibinfo{author}{\bibfnamefont{P.}~\bibnamefont{Balcou}},
  \bibinfo{author}{\bibfnamefont{K.~J.} \bibnamefont{Schafer}},
  \bibinfo{author}{\bibfnamefont{K.~C.} \bibnamefont{Kulander}},
  \bibnamefont{and}
  \bibinfo{author}{\bibfnamefont{A.}~\bibnamefont{L'Huillier}},
  \bibinfo{journal}{Phys. Rev. A} \textbf{\bibinfo{volume}{59}},
  \bibinfo{pages}{1367} (\bibinfo{year}{1999}),
  \urlprefix\url{https://link.aps.org/doi/10.1103/PhysRevA.59.1367}.

\bibitem[{\citenamefont{Sali\`eres et~al.}(1998)\citenamefont{Sali\`eres,
  Antoine, de~Bohan, and Lewenstein}}]{SaAnPh98}
\bibinfo{author}{\bibfnamefont{P.}~\bibnamefont{Sali\`eres}},
  \bibinfo{author}{\bibfnamefont{P.}~\bibnamefont{Antoine}},
  \bibinfo{author}{\bibfnamefont{A.}~\bibnamefont{de~Bohan}}, \bibnamefont{and}
  \bibinfo{author}{\bibfnamefont{M.}~\bibnamefont{Lewenstein}},
  \bibinfo{journal}{Phys. Rev. Lett.} \textbf{\bibinfo{volume}{81}},
  \bibinfo{pages}{5544} (\bibinfo{year}{1998}),
  \urlprefix\url{https://link.aps.org/doi/10.1103/PhysRevLett.81.5544}.

\bibitem[{\citenamefont{Antoine et~al.}(1997)\citenamefont{Antoine,
  Milo\ifmmode \check{s}\else \v{s}\fi{}evi\ifmmode~\acute{c}\else \'{c}\fi{},
  L'Huillier, Gaarde, Sali\`eres, and Lewenstein}}]{AnMiDe97}
\bibinfo{author}{\bibfnamefont{P.}~\bibnamefont{Antoine}},
  \bibinfo{author}{\bibfnamefont{D.~B.} \bibnamefont{Milo\ifmmode
  \check{s}\else \v{s}\fi{}evi\ifmmode~\acute{c}\else \'{c}\fi{}}},
  \bibinfo{author}{\bibfnamefont{A.}~\bibnamefont{L'Huillier}},
  \bibinfo{author}{\bibfnamefont{M.~B.} \bibnamefont{Gaarde}},
  \bibinfo{author}{\bibfnamefont{P.}~\bibnamefont{Sali\`eres}},
  \bibnamefont{and}
  \bibinfo{author}{\bibfnamefont{M.}~\bibnamefont{Lewenstein}},
  \bibinfo{journal}{Phys. Rev. A} \textbf{\bibinfo{volume}{56}},
  \bibinfo{pages}{4960} (\bibinfo{year}{1997}),
  \urlprefix\url{https://link.aps.org/doi/10.1103/PhysRevA.56.4960}.

\bibitem[{\citenamefont{Takahashi et~al.}(2007)\citenamefont{Takahashi, Kanai,
  Ishikawa, Nabekawa, and Midorikawa}}]{TaKaIs07}
\bibinfo{author}{\bibfnamefont{E.~J.} \bibnamefont{Takahashi}},
  \bibinfo{author}{\bibfnamefont{T.}~\bibnamefont{Kanai}},
  \bibinfo{author}{\bibfnamefont{K.~L.} \bibnamefont{Ishikawa}},
  \bibinfo{author}{\bibfnamefont{Y.}~\bibnamefont{Nabekawa}}, \bibnamefont{and}
  \bibinfo{author}{\bibfnamefont{K.}~\bibnamefont{Midorikawa}},
  \bibinfo{journal}{Phys. Rev. Lett.} \textbf{\bibinfo{volume}{99}},
  \bibinfo{pages}{053904} (\bibinfo{year}{2007}),
  \urlprefix\url{https://link.aps.org/doi/10.1103/PhysRevLett.99.053904}.

\bibitem[{\citenamefont{Husakou et~al.}(2011)\citenamefont{Husakou, Im, and
  Herrmann}}]{plasmonHusakou2011}
\bibinfo{author}{\bibfnamefont{A.}~\bibnamefont{Husakou}},
  \bibinfo{author}{\bibfnamefont{S.-J.} \bibnamefont{Im}}, \bibnamefont{and}
  \bibinfo{author}{\bibfnamefont{J.}~\bibnamefont{Herrmann}},
  \bibinfo{journal}{Physical Review A} \textbf{\bibinfo{volume}{83}},
  \bibinfo{pages}{043839} (\bibinfo{year}{2011}), ISSN
  \bibinfo{issn}{1050-2947}, \eprint{1009.4124},
  \urlprefix\url{http://link.aps.org/doi/10.1103/PhysRevA.83.043839}.

\bibitem[{\citenamefont{Kim et~al.}(2008)\citenamefont{Kim, Jin, Kim, Park,
  Kim, and Kim}}]{plasmonKim2008}
\bibinfo{author}{\bibfnamefont{S.}~\bibnamefont{Kim}},
  \bibinfo{author}{\bibfnamefont{J.}~\bibnamefont{Jin}},
  \bibinfo{author}{\bibfnamefont{Y.-J.} \bibnamefont{Kim}},
  \bibinfo{author}{\bibfnamefont{I.-Y.} \bibnamefont{Park}},
  \bibinfo{author}{\bibfnamefont{Y.}~\bibnamefont{Kim}}, \bibnamefont{and}
  \bibinfo{author}{\bibfnamefont{S.-W.} \bibnamefont{Kim}},
  \bibinfo{journal}{Nature} \textbf{\bibinfo{volume}{453}},
  \bibinfo{pages}{757} (\bibinfo{year}{2008}), ISSN \bibinfo{issn}{0028-0836}.

\bibitem[{\citenamefont{Yavuz et~al.}(2012)\citenamefont{Yavuz, Bleda, Altun,
  and Topcu}}]{inhomogYavuz2012}
\bibinfo{author}{\bibfnamefont{I.}~\bibnamefont{Yavuz}},
  \bibinfo{author}{\bibfnamefont{E.~A.} \bibnamefont{Bleda}},
  \bibinfo{author}{\bibfnamefont{Z.}~\bibnamefont{Altun}}, \bibnamefont{and}
  \bibinfo{author}{\bibfnamefont{T.}~\bibnamefont{Topcu}},
  \bibinfo{journal}{Physical Review A} \textbf{\bibinfo{volume}{85}},
  \bibinfo{pages}{013416} (\bibinfo{year}{2012}), ISSN
  \bibinfo{issn}{1050-2947},
  \urlprefix\url{http://link.aps.org/doi/10.1103/PhysRevA.85.013416}.

\bibitem[{\citenamefont{Ciappina et~al.}(2012)\citenamefont{Ciappina, Biegert,
  Quidant, and Lewenstein}}]{inhomogCiappina2012}
\bibinfo{author}{\bibfnamefont{M.~F.} \bibnamefont{Ciappina}},
  \bibinfo{author}{\bibfnamefont{J.}~\bibnamefont{Biegert}},
  \bibinfo{author}{\bibfnamefont{R.}~\bibnamefont{Quidant}}, \bibnamefont{and}
  \bibinfo{author}{\bibfnamefont{M.}~\bibnamefont{Lewenstein}},
  \bibinfo{journal}{Physical Review A} \textbf{\bibinfo{volume}{85}},
  \bibinfo{pages}{033828} (\bibinfo{year}{2012}), ISSN
  \bibinfo{issn}{1050-2947},
  \urlprefix\url{http://link.aps.org/doi/10.1103/PhysRevA.85.033828}.

\bibitem[{\citenamefont{Corkum}(1993)}]{Corkum1993}
\bibinfo{author}{\bibfnamefont{P.~B.} \bibnamefont{Corkum}},
  \bibinfo{journal}{Physical Review Letters} \textbf{\bibinfo{volume}{71}},
  \bibinfo{pages}{1994} (\bibinfo{year}{1993}).

\bibitem[{\citenamefont{Topcu and Robicheaux}(2012)}]{TopRob12}
\bibinfo{author}{\bibfnamefont{T.}~\bibnamefont{Topcu}} \bibnamefont{and}
  \bibinfo{author}{\bibfnamefont{F.}~\bibnamefont{Robicheaux}},
  \bibinfo{journal}{Physical Review A} \textbf{\bibinfo{volume}{86}},
  \bibinfo{pages}{053407} (\bibinfo{year}{2012}), ISSN
  \bibinfo{issn}{1050-2947},
  \urlprefix\url{http://link.aps.org/doi/10.1103/PhysRevA.86.053407}.

\bibitem[{\citenamefont{Bleda et~al.}(2013)\citenamefont{Bleda, Yavuz, Altun,
  and Topcu}}]{BleYavAlt13}
\bibinfo{author}{\bibfnamefont{E.~A.} \bibnamefont{Bleda}},
  \bibinfo{author}{\bibfnamefont{I.}~\bibnamefont{Yavuz}},
  \bibinfo{author}{\bibfnamefont{Z.}~\bibnamefont{Altun}}, \bibnamefont{and}
  \bibinfo{author}{\bibfnamefont{T.}~\bibnamefont{Topcu}},
  \bibinfo{journal}{Physical Review A} \textbf{\bibinfo{volume}{88}},
  \bibinfo{pages}{043417} (\bibinfo{year}{2013}), ISSN
  \bibinfo{issn}{1050-2947},
  \urlprefix\url{http://link.aps.org/doi/10.1103/PhysRevA.88.043417}.

\bibitem[{\citenamefont{Top{\c{c}}u and Robicheaux}(2007)}]{topcu07}
\bibinfo{author}{\bibfnamefont{T.}~\bibnamefont{Top{\c{c}}u}} \bibnamefont{and}
  \bibinfo{author}{\bibfnamefont{F.}~\bibnamefont{Robicheaux}},
  \bibinfo{journal}{Journal of Physics B: Atomic, Molecular and Optical
  Physics} \textbf{\bibinfo{volume}{40}}, \bibinfo{pages}{1925}
  (\bibinfo{year}{2007}), ISSN \bibinfo{issn}{0953-4075},
  \urlprefix\url{http://stacks.iop.org/0953-4075/40/i=10/a=025?key=crossref.d4188e63f0e9fd29c18fd412297f0d85}.

\bibitem[{\citenamefont{Jaro{\'{n}}-Becker
  et~al.}(2006)\citenamefont{Jaro{\'{n}}-Becker, Becker, and
  Faisal}}]{BecBecFai06}
\bibinfo{author}{\bibfnamefont{A.}~\bibnamefont{Jaro{\'{n}}-Becker}},
  \bibinfo{author}{\bibfnamefont{A.}~\bibnamefont{Becker}}, \bibnamefont{and}
  \bibinfo{author}{\bibfnamefont{F.~H.} \bibnamefont{Faisal}},
  \bibinfo{journal}{Physical Review Letters} \textbf{\bibinfo{volume}{96}},
  \bibinfo{pages}{2} (\bibinfo{year}{2006}), ISSN \bibinfo{issn}{00319007}.

\bibitem[{\citenamefont{Hunsche et~al.}(1996)\citenamefont{Hunsche,
  Starczewski, l'Huillier, Persson, Wahlstr\"om, van Linden van~den Heuvell,
  and Svanberg}}]{HunStaHui96}
\bibinfo{author}{\bibfnamefont{S.}~\bibnamefont{Hunsche}},
  \bibinfo{author}{\bibfnamefont{T.}~\bibnamefont{Starczewski}},
  \bibinfo{author}{\bibfnamefont{A.}~\bibnamefont{l'Huillier}},
  \bibinfo{author}{\bibfnamefont{A.}~\bibnamefont{Persson}},
  \bibinfo{author}{\bibfnamefont{C.-G.} \bibnamefont{Wahlstr\"om}},
  \bibinfo{author}{\bibfnamefont{H.~B.} \bibnamefont{van Linden van~den
  Heuvell}}, \bibnamefont{and}
  \bibinfo{author}{\bibfnamefont{S.}~\bibnamefont{Svanberg}},
  \bibinfo{journal}{Phys. Rev. Lett.} \textbf{\bibinfo{volume}{77}},
  \bibinfo{pages}{1966} (\bibinfo{year}{1996}),
  \urlprefix\url{https://link.aps.org/doi/10.1103/PhysRevLett.77.1966}.

\bibitem[{\citenamefont{Ciappina et~al.}(2008)\citenamefont{Ciappina, Becker,
  and Jaro{\'{n}}-Becker}}]{CiaBecBec08}
\bibinfo{author}{\bibfnamefont{M.~F.} \bibnamefont{Ciappina}},
  \bibinfo{author}{\bibfnamefont{A.}~\bibnamefont{Becker}}, \bibnamefont{and}
  \bibinfo{author}{\bibfnamefont{A.}~\bibnamefont{Jaro{\'{n}}-Becker}},
  \bibinfo{journal}{Physical Review A - Atomic, Molecular, and Optical Physics}
  \textbf{\bibinfo{volume}{78}}, \bibinfo{pages}{1} (\bibinfo{year}{2008}),
  ISSN \bibinfo{issn}{10502947}.

\bibitem[{\citenamefont{Dolmatov et~al.}(2009)\citenamefont{Dolmatov, Craven,
  Guler, and Keating}}]{DolCraGul09}
\bibinfo{author}{\bibfnamefont{V.~K.} \bibnamefont{Dolmatov}},
  \bibinfo{author}{\bibfnamefont{G.~T.} \bibnamefont{Craven}},
  \bibinfo{author}{\bibfnamefont{E.}~\bibnamefont{Guler}}, \bibnamefont{and}
  \bibinfo{author}{\bibfnamefont{D.}~\bibnamefont{Keating}},
  \bibinfo{journal}{Physical Review A - Atomic, Molecular, and Optical Physics}
  \textbf{\bibinfo{volume}{80}}, \bibinfo{pages}{1} (\bibinfo{year}{2009}),
  ISSN \bibinfo{issn}{10502947}, \eprint{0908.2241}.

\bibitem[{\citenamefont{Müller et~al.}(2007)\citenamefont{Müller, Schippers,
  Phaneuf, Habibi, Esteves, Wang, Kilcoyne, Aguilar, Yang, and
  Dunsch}}]{MuScPh07}
\bibinfo{author}{\bibfnamefont{A.}~\bibnamefont{Müller}},
  \bibinfo{author}{\bibfnamefont{S.}~\bibnamefont{Schippers}},
  \bibinfo{author}{\bibfnamefont{R.~A.} \bibnamefont{Phaneuf}},
  \bibinfo{author}{\bibfnamefont{M.}~\bibnamefont{Habibi}},
  \bibinfo{author}{\bibfnamefont{D.}~\bibnamefont{Esteves}},
  \bibinfo{author}{\bibfnamefont{J.~C.} \bibnamefont{Wang}},
  \bibinfo{author}{\bibfnamefont{A.~L.~D.} \bibnamefont{Kilcoyne}},
  \bibinfo{author}{\bibfnamefont{A.}~\bibnamefont{Aguilar}},
  \bibinfo{author}{\bibfnamefont{S.}~\bibnamefont{Yang}}, \bibnamefont{and}
  \bibinfo{author}{\bibfnamefont{L.}~\bibnamefont{Dunsch}},
  \bibinfo{journal}{Journal of Physics: Conference Series}
  \textbf{\bibinfo{volume}{88}}, \bibinfo{pages}{012038}
  (\bibinfo{year}{2007}),
  \urlprefix\url{http://stacks.iop.org/1742-6596/88/i=1/a=012038}.

\bibitem[{\citenamefont{Chen et~al.}(2010)\citenamefont{Chen, Phaneuf, and
  Msezane}}]{ChPhMs10}
\bibinfo{author}{\bibfnamefont{Z.}~\bibnamefont{Chen}},
  \bibinfo{author}{\bibfnamefont{R.~A.} \bibnamefont{Phaneuf}},
  \bibnamefont{and} \bibinfo{author}{\bibfnamefont{A.~Z.}
  \bibnamefont{Msezane}}, \bibinfo{journal}{Journal of Physics B: Atomic,
  Molecular and Optical Physics} \textbf{\bibinfo{volume}{43}},
  \bibinfo{pages}{215203} (\bibinfo{year}{2010}),
  \urlprefix\url{http://stacks.iop.org/0953-4075/43/i=21/a=215203}.

\bibitem[{\citenamefont{Ni et~al.}(2008)\citenamefont{Ni, Zamith, L\'epine,
  Martchenko, Kling, Ghafur, Muller, Berden, Robicheaux, and
  Vrakking}}]{vrakking}
\bibinfo{author}{\bibfnamefont{Y.}~\bibnamefont{Ni}},
  \bibinfo{author}{\bibfnamefont{S.}~\bibnamefont{Zamith}},
  \bibinfo{author}{\bibfnamefont{F.}~\bibnamefont{L\'epine}},
  \bibinfo{author}{\bibfnamefont{T.}~\bibnamefont{Martchenko}},
  \bibinfo{author}{\bibfnamefont{M.}~\bibnamefont{Kling}},
  \bibinfo{author}{\bibfnamefont{O.}~\bibnamefont{Ghafur}},
  \bibinfo{author}{\bibfnamefont{H.~G.} \bibnamefont{Muller}},
  \bibinfo{author}{\bibfnamefont{G.}~\bibnamefont{Berden}},
  \bibinfo{author}{\bibfnamefont{F.}~\bibnamefont{Robicheaux}},
  \bibnamefont{and} \bibinfo{author}{\bibfnamefont{M.~J.~J.}
  \bibnamefont{Vrakking}}, \bibinfo{journal}{Phys. Rev. A}
  \textbf{\bibinfo{volume}{78}}, \bibinfo{pages}{013413}
  (\bibinfo{year}{2008}),
  \urlprefix\url{https://link.aps.org/doi/10.1103/PhysRevA.78.013413}.

\bibitem[{\citenamefont{Bandrauk et~al.}(2009)\citenamefont{Bandrauk,
  Chelkowski, Diestler, Manz, and Yuan}}]{BanCheDie09}
\bibinfo{author}{\bibfnamefont{A.~D.} \bibnamefont{Bandrauk}},
  \bibinfo{author}{\bibfnamefont{S.}~\bibnamefont{Chelkowski}},
  \bibinfo{author}{\bibfnamefont{D.~J.} \bibnamefont{Diestler}},
  \bibinfo{author}{\bibfnamefont{J.}~\bibnamefont{Manz}}, \bibnamefont{and}
  \bibinfo{author}{\bibfnamefont{K.~J.} \bibnamefont{Yuan}},
  \bibinfo{journal}{Physical Review A - Atomic, Molecular, and Optical Physics}
  \textbf{\bibinfo{volume}{79}}, \bibinfo{pages}{1} (\bibinfo{year}{2009}),
  ISSN \bibinfo{issn}{10502947}.

\bibitem[{\citenamefont{Reittu}(1995)}]{Rei95}
\bibinfo{author}{\bibfnamefont{H.~J.} \bibnamefont{Reittu}},
  \bibinfo{journal}{American Journal of Physics} \textbf{\bibinfo{volume}{63}},
  \bibinfo{pages}{940} (\bibinfo{year}{1995}),
  \eprint{https://doi.org/10.1119/1.18037},
  \urlprefix\url{https://doi.org/10.1119/1.18037}.

\bibitem[{\citenamefont{Zhai et~al.}(2011)\citenamefont{Zhai, Zhu, Chen, Yan,
  Fu, and Wang}}]{ZhZhCh11}
\bibinfo{author}{\bibfnamefont{Z.}~\bibnamefont{Zhai}},
  \bibinfo{author}{\bibfnamefont{Q.}~\bibnamefont{Zhu}},
  \bibinfo{author}{\bibfnamefont{J.}~\bibnamefont{Chen}},
  \bibinfo{author}{\bibfnamefont{Z.-C.} \bibnamefont{Yan}},
  \bibinfo{author}{\bibfnamefont{P.}~\bibnamefont{Fu}}, \bibnamefont{and}
  \bibinfo{author}{\bibfnamefont{B.}~\bibnamefont{Wang}},
  \bibinfo{journal}{Phys. Rev. A} \textbf{\bibinfo{volume}{83}},
  \bibinfo{pages}{043409} (\bibinfo{year}{2011}),
  \urlprefix\url{https://link.aps.org/doi/10.1103/PhysRevA.83.043409}.

\bibitem[{\citenamefont{Zhai et~al.}(2010)\citenamefont{Zhai, Chen, Yan, Fu,
  and Wang}}]{ZhChYa10}
\bibinfo{author}{\bibfnamefont{Z.}~\bibnamefont{Zhai}},
  \bibinfo{author}{\bibfnamefont{J.}~\bibnamefont{Chen}},
  \bibinfo{author}{\bibfnamefont{Z.-C.} \bibnamefont{Yan}},
  \bibinfo{author}{\bibfnamefont{P.}~\bibnamefont{Fu}}, \bibnamefont{and}
  \bibinfo{author}{\bibfnamefont{B.}~\bibnamefont{Wang}},
  \bibinfo{journal}{Phys. Rev. A} \textbf{\bibinfo{volume}{82}},
  \bibinfo{pages}{043422} (\bibinfo{year}{2010}),
  \urlprefix\url{https://link.aps.org/doi/10.1103/PhysRevA.82.043422}.

\end{thebibliography}
	
	\end{document}